\documentclass[10pt]{article}
\usepackage[cp1251]{inputenc}
\usepackage[dvips]{epsfig}
\usepackage[dvips]{changebar}
\usepackage[dvips]{graphicx}
\usepackage{amsmath,enumerate, amsfonts, amssymb,amsthm}
\usepackage{cite}

\def\lsim{\
  \lower-1.2pt\vbox{\hbox{\rlap{$<$}\lower5pt\vbox{\hbox{$\sim$}}}}\ }
\def\gsim{\
  \lower-1.2pt\vbox{\hbox{\rlap{$>$}\lower5pt\vbox{\hbox{$\sim$}}}}\ }

\begin{document}

\title{Nonuniform Bose-Einstein condensate. I.\\
 An improvement of the Gross-Pitaevskii method}
\author{Maksim Tomchenko \bigskip \\
{\small Bogolyubov Institute for Theoretical Physics} \\
{\small 14b, Metrolohichna Str., Kyiv 03143, Ukraine}\\
{\small E-mail:mtomchenko@bitp.kyiv.ua}}
\date{\empty }
\maketitle

\textit{A nonuniform condensate is usually described by the
Gross-Pitaevskii (GP) equation, which is derived with the help of
the c-number ansatz $\hat{ \Psi}(\mathbf{r},t)=\Psi (\mathbf{r},t)$.
Proceeding from a more accurate operator ansatz
$\hat{\Psi}(\mathbf{r},t)=\hat{a}_{0}\Psi (\mathbf{r},t)/ \sqrt{N}$,
where $N$ is the number of Bose particles, we find the equation
$i\hbar \frac{\partial \Psi (\mathbf{r},t)}{\partial t}=-\frac{\hbar
^{2}}{2m}\frac{\partial ^{2}\Psi (\mathbf{r},t)}{\partial
\mathbf{r}^{2}}+\left( 1-\frac{1}{N}\right) 2c\Psi
(\mathbf{r},t)|\Psi(\mathbf{r},t)|^{2}$, which we call the GP$_{N}$
equation. It differs from the GP equation by the factor $\left(
1-1/N\right) $. We compare the accuracy of the GP and GP$_{N}$
equations by analysing the ground state of a one-dimensional system
of point bosons with repulsive interaction ($c>0$) and zero boundary
conditions. Both equations are solved numerically, and the system
energy $E$ and the particle density profile $\rho (x)$ are
determined for the mean particle density $\bar{\rho}=1$, different
values of $N$ and of the coupling constant $\gamma =c/\bar{\rho}$.
The solutions are compared with the exact ones obtained by the Bethe
ansatz. The results show that the GP and GP$_{N}$ equations equally
well describe the many-particle system ($N\gtrsim 100$) being in the
weak coupling regime ($N^{-2}\ll \gamma \lesssim 0.1$). But for the
few-boson system ($N\lesssim 10$) with $\gamma \lesssim N^{-2}$ the
solutions of the GP$_{N}$ equation are in much better  agreement
with the exact ones. Thus, the multiplier $\left( 1-1/N \right) $
allows one to describe few-boson systems with high accuracy. This
means that it is reasonable to extend the notion of Bose-Einstein
condensation to few-particle systems.}\\
\textbf{Keywords:} nonuniform condensate;  the Gross-Pitaevskii
equation; ground state.

\section{Introduction}

The simplest analytical method for describing a system of $N$
spinless interacting bosons is based on the solution of the
nonlinear Schr\"{o}dinger
equation%
\begin{equation}
i\hbar \frac{\partial \Psi (\mathbf{r},t)}{\partial t}=-\frac{\hbar ^{2}}{2m}%
\frac{\partial ^{2}\Psi (\mathbf{r},t)}{\partial \mathbf{r}^{2}}+\Psi (%
\mathbf{r},t)\int\limits_{V}d\mathbf{r}^{\prime }U(|\mathbf{r}-\mathbf{r}%
^{\prime }|)|\Psi (\mathbf{r}^{\prime },t)|^{2},  \label{1}
\end{equation}%
where $\Psi $ is the wave function of condensate,  $V$ is the system
volume. This equation was first obtained by E.~Gross in 1958~\cite%
{gross1957,gross1958}. Gross perceived that N.~Bogoliubov's approach~\cite%
{bog1947} could be applied to describe a \textit{nonuniform}
condensate if we set $\hat{\Psi}(\mathbf{r},t)=\Psi (\mathbf{r},t)$.
In this case, the Heisenberg equation becomes Eq.~(\ref{1}) for
$\Psi (\mathbf{r},t)$. A few years later, Eq.~(\ref{1}) was written
by L. Pitaevskii~\cite{pit1961}
and E.~Gross \cite{gross1961} for the point potential $U(|\mathbf{r}_{j}-%
\mathbf{r}_{l}|)=2c\delta (\mathbf{r}_{j}-\mathbf{r}_{l})$,%
\begin{equation}
i\hbar \frac{\partial \Psi (\mathbf{r},t)}{\partial t}=-\frac{\hbar ^{2}}{2m}%
\frac{\partial ^{2}\Psi (\mathbf{r},t)}{\partial \mathbf{r}^{2}}+2c\Psi (%
\mathbf{r},t)|\Psi (\mathbf{r},t)|^{2}.  \label{2}
\end{equation}%
Since real-world atoms have a non-zero size, it is more accurate to
describe them by Eq.~(\ref{1}). However, if the characteristic
dimensions of inhomogeneities in the system are much larger than the
atomic size, the
Gross-Pitaevskii (GP) equation~(\ref{2}) may be used instead of Eq.~(\ref{1}%
). The GP equation is simpler than the Gross equation~(\ref{1}) and
is basic for describing the dilute Bose gas in a trap. A huge number
of experimental and theoretical works published during the last 25
years were
devoted to the study of gases in the trap~\cite%
{leggett2001,pethick2008,blume2012,pitstring}. The Nobel Prize was
awarded for
the experimental production of Bose condensate~\cite%
{cornell2002,ketterle2002}.

It is generally accepted that Eqs.~(\ref{1}) and (\ref{2}) give a
semiclassical description of the system and are only applicable to
systems with a large number $N$ of particles. Several arguments have
been made in favour of
this~\cite{pitstring,fetter,pinesnoz,lieb2005}. The main one is that
the Bogoliubov method works namely at $N\gg 1$.

It is often asserted that Eqs.~(\ref{1}) and (\ref{2}) can also be
derived from the condensate approximation for the total $N$-particle
wave function of the system
\begin{equation}
\Psi _{N}(\mathbf{r}_{1},\ldots ,\mathbf{r}_{N},t)=\prod\limits_{j=1}^{N}%
\psi (\mathbf{r}_{j},t).  \label{3}
\end{equation}%
However, this is not quite so.  By integrating the $N$-particle
Schr\"{o}dinger equation \cite{esryphd} and by using a variational
approach \cite{salasnich2000} it was shown that ansatz (\ref{3})
gives rise to the equation
\begin{eqnarray}
&&i\hbar \frac{\partial \Psi (\mathbf{r},t)}{\partial
t}=-\frac{\hbar ^{2}}{2m}\frac{\partial ^{2}\Psi
(\mathbf{r},t)}{\partial \mathbf{r}^{2}} +\left( 1-\frac{1}{N}\right) \Psi (\mathbf{r},t)\int\limits_{V}d\mathbf{r}%
^{\prime }U(|\mathbf{r}-\mathbf{r}^{\prime }|)|\Psi
(\mathbf{r}^{\prime },t)|^{2},  \label{4}
\end{eqnarray}%
where $\Psi (\mathbf{r},t)= e^{-i\kappa(t)/\hbar}\sqrt{N}\psi
(\mathbf{r},t)$ (see section 2.1).
For the potential $U(|\mathbf{r}_{j}-\mathbf{r}_{l}|)=2c\delta (\mathbf{r}_{j}-\mathbf{r}%
_{l})$ it transforms into
\begin{equation}
i\hbar \frac{\partial \Psi (\mathbf{r},t)}{\partial t}=-\frac{\hbar
^{2}}{2m}\frac{\partial ^{2}\Psi (\mathbf{r},t)}{\partial
\mathbf{r}^{2}}+\left( 1-\frac{1}{N}\right) 2c\Psi (\mathbf{r},t)|\Psi (%
\mathbf{r},t)|^{2}.  \label{6}
\end{equation}%
It will be seen in section~2 below that if the more accurate
operator ansatz
\begin{equation}
\hat{\Psi}(\mathbf{r},t)=\hat{a}_{0}\Psi (\mathbf{r},t)/\sqrt{N}  \label{5}
\end{equation}%
is used instead of the c-number ansatz $\hat{\Psi}(\mathbf{r},t)=\Psi (%
\mathbf{r},t)$, we also get Eqs.~(\ref{4}) and (\ref{6}).

\textit{Equations (\ref{2}) and (\ref{6}) will be called the GP and
GP$_{N}$ equations, respectively.} The GP$_{N}$ equation differs
from the GP one by the factor $\left( 1-1/N\right) $. This equation
can be obtained using  ans\"{a}tze (\ref{3}) and (\ref{5}), which are
valid for any $N\geq 2$ (in contrast to the ansatz $\hat{\Psi%
}(\mathbf{r},t)\approx \Psi (\mathbf{r},t)$, which is only applicable for $%
N\gg 1$). This indicates that the GP$_{N}$ equation must be able to
describe a Bose system even for small $N$.

The accuracy of the GP$_{N}$ equation has already been investigated
for a Bose gas under spherically symmetric harmonic confinement by
comparing solutions of the GP$_{N}$ equation with Monte Carlo
numerical solutions for $N=2$--$50$ \cite{blume2001} and with
analytical solutions of the linear Schr\"{o}dinger equation for
$N=2$ \cite{blume2012}. Such an analysis showed that for $0\leq
(N-1)a/a_{h0}\lsim 0.1$ (where $a$ is the s-wave scattering length,
and $a_{h0}=(\hbar/m\omega_{h0})^{1/2}$), the GP$_{N}$ equation
describes the system with very good accuracy. However, the
comparison of the accuracy of the GP$_{N}$ and GP equations has not
yet been carried out.

In this and the next paper~\cite{gp2} we study in detail a
one-dimensional (1D) Bose gas in the absence of a trap and compare
the solutions of  stationary GP and GP$_{N}$ equations for different
$N\geq 2$ with the exact Bethe-ansatz solutions. In this paper, an
equation for a nonuniform condensate is derived (section~2) and its
solutions for the ground state of the condensate are analysed
(sections~3--5). In the next article \cite{gp2}, the excited  states
of the condensate are considered.

Note that we became aware of papers
\cite{esryphd,salasnich2000,blume2001,blume2012} after this article
was submitted to arXiv.  Moreover, after this article and \cite{gp2}
had already been written, we learned that analogous solutions of the
GP equation have been found analytically by L.~Carr, C.~Clark, and
W.~Reinhardt~\cite{carr2000r}. Their solutions are expressed in
terms of the Jacobi elliptic functions. In this paper, we obtain
solutions using a different (numerical) method. Our solutions are
consistent with those of work~\cite{carr2000r}, although we have not
made a detailed one-to-one comparison.

For a 1D system, the term quasi-condensate is usually used instead
of condensate~\cite{pethick2008,petrov2004,bouchoule2009}. For
brevity, we will omit the prefix \textquotedblleft
quasi\textquotedblright\ and write \textquotedblleft
condensate\textquotedblright, even for small $N$.

\section{Derivation of equation for nonuniform condensate}

\subsection{Wave-function approach}
This approach was first proposed, apparently by B.~Esry
\cite{esryphd}, but it has not been published.
Consider a system of $N\geq 2$ interacting spinless bosons. The Schr%
\"{o}dinger equation reads
\begin{equation}
i\hbar \frac{\partial \Psi _{N}}{\partial t}=-\frac{\hbar ^{2}}{2m}%
\sum\limits_{j=1}^{N}\frac{\partial ^{2}}{\partial \mathbf{r}_{j}^{2}}\Psi
_{N}+\frac{1}{2}\mathop{\sum^{N}_{j,p=1}}\limits_{(j\neq p)}U(|\mathbf{r}%
_{p}-\mathbf{r}_{j}|)\Psi _{N}.  \label{p1-1}
\end{equation}%
Let the wave function $\Psi _{N}(\mathbf{r}_{1},\ldots
,\mathbf{r}_{N},t)$ of the system have the condensate form (\ref{3})
with the normalisation $\int_{V}d\mathbf{r}|\psi
(\mathbf{r},t)|^{2}=1$. Substituting (\ref{3}) into (\ref{p1-1}), we
obtain
\begin{eqnarray}
&&i\hbar \sum\limits_{j=1}^{N}\mathop{\prod^{N}_{l=1}}\limits_{(l\neq
j)}\psi (\mathbf{r}_{l},t)\cdot \frac{\partial \psi (\mathbf{r}_{j},t)}{%
\partial t}=  \notag \\
&=&-\frac{\hbar ^{2}}{2m}\sum\limits_{j=1}^{N}\mathop{\prod^{N}_{l=1}}%
\limits_{(l\neq j)}\psi (\mathbf{r}_{l},t)\cdot \frac{\partial ^{2}\psi (%
\mathbf{r}_{j},t)}{\partial \mathbf{r}_{j}^{2}}+\frac{1}{2}%
\mathop{\sum^{N}_{j,p=1}}\limits_{(p\neq j)}U(|\mathbf{r}_{p}-\mathbf{r}%
_{j}|)\prod\limits_{l=1}^{N}\psi (\mathbf{r}_{l},t).  \label{p1-2}
\end{eqnarray}%
Multiplying this equation by $\prod_{l=2}^{N}\psi ^{\ast }(\mathbf{r}_{l},t)$
and integrating the result over $\mathbf{r}_{2},\ldots ,\mathbf{r}_{N}$, we
get
\begin{eqnarray}
&&i\hbar \frac{\partial \psi (\mathbf{r}_{1},t)}{\partial t}-a(t)\psi (%
\mathbf{r}_{1},t)=-\frac{\hbar ^{2}}{2m}\frac{\partial ^{2}\psi (\mathbf{r}%
_{1},t)}{\partial \mathbf{r}_{1}^{2}}  \notag \\
&&+(N-1)\psi (\mathbf{r}_{1},t)\int\limits_{V}d\mathbf{r}_{2}U(|\mathbf{r}%
_{1}-\mathbf{r}_{2}|)|\psi (\mathbf{r}_{2},t)|^{2},  \label{p1-3}
\end{eqnarray}%
where%
\begin{eqnarray}
a(t) &=&(N-1)\int\limits_{V}d\mathbf{r}_{2}\psi ^{\ast }(\mathbf{r}%
_{2},t)\left\{ -\frac{\hbar ^{2}}{2m}\frac{\partial ^{2}\psi (\mathbf{r}%
_{2},t)}{\partial \mathbf{r}_{2}^{2}}-i\hbar \frac{\partial \psi (\mathbf{r}%
_{2},t)}{\partial t}\right\}  \notag \\
&+&\frac{(N-1)(N-2)}{2}\int\limits_{V}d\mathbf{r}_{2}d\mathbf{r}_{3}U(|%
\mathbf{r}_{2}-\mathbf{r}_{3}|)|\psi (\mathbf{r}_{2},t)|^{2}|\psi (\mathbf{r}%
_{3},t)|^{2}.  \label{p1-4}
\end{eqnarray}%
The derivatives in the right-hand side of (\ref{p1-4}) can be
expressed using Eq.~(\ref{p1-3}), from whence
\begin{equation}
a(t)=-\frac{(N-1)}{2}\int\limits_{V}d\mathbf{r}_{2}d\mathbf{r}_{3}U(|\mathbf{%
r}_{2}-\mathbf{r}_{3}|)|\psi (\mathbf{r}_{2},t)|^{2}|\psi (\mathbf{r}%
_{3},t)|^{2}.  \label{p1-5}
\end{equation}%
Let us set
$\psi(\mathbf{r}_{1},t)=e^{i\kappa(t)/\hbar}\tilde{\psi}(\mathbf{r}_{1},t)$,
where $\kappa(t)=-\int_{-\infty}^{t}d\tau a(\tau)$. Then
(\ref{p1-3}) is reduced to \cite{esryphd}
\begin{eqnarray}
 i\hbar\frac{\partial \tilde{\psi}(\mathbf{r}_{1},t)}{\partial t} =
-\frac{\hbar^2}{2m}\frac{\partial^{2}
\tilde{\psi}(\mathbf{r}_{1},t)}{\partial \mathbf{r}_{1}^{2}} +
(N-1)\tilde{\psi}(\mathbf{r}_{1},t) \int\limits_{V} d \mathbf{r}_{2}
 U(|\mathbf{r}_{1}-\mathbf{r}_{2}|)|\tilde{\psi}(\mathbf{r}_{2},t)|^{2}.
     \label{p1-7} \end{eqnarray}
Making the substitution $\tilde{\psi} (\mathbf{r},t)=\Psi
(\mathbf{r},t)/\sqrt{N}$, we obtain the final equation
\begin{eqnarray}
&&i\hbar \frac{\partial \Psi (\mathbf{r},t)}{\partial
t}=-\frac{\hbar ^{2}}{2m}\frac{\partial ^{2}\Psi
(\mathbf{r},t)}{\partial
\mathbf{r}^{2}}
+\left( 1-\frac{1}{N}\right) \Psi (\mathbf{r},t)\int\limits_{V}d\mathbf{r}%
^{\prime }U(|\mathbf{r}-\mathbf{r}^{\prime }|)|\Psi (\mathbf{r}^{\prime
},t)|^{2},  \label{p1-6}
\end{eqnarray}%
with the normalisation $\int_{V}d\mathbf{r}|\Psi
(\mathbf{r},t)|^{2}=N$. For
the point potential $U(|\mathbf{r}_{j}-\mathbf{r}_{l}|)=2c\delta (\mathbf{r}%
_{j}-\mathbf{r}_{l})$, Eq. (\ref{p1-6}) reads
\begin{equation}
i\hbar \frac{\partial \Psi (\mathbf{r},t)}{\partial t}=-\frac{\hbar
^{2}}{2m}\frac{\partial ^{2}\Psi (\mathbf{r},t)}{\partial
\mathbf{r}^{2}}+\left( 1-\frac{1}{N}\right) 2c\Psi (\mathbf{r},t)|\Psi (%
\mathbf{r},t)|^{2}.  \label{p1-8}
\end{equation}%

It is worth noting that the condensate ansatz (\ref{3}) does not
satisfy the Schr\"{o}dinger equation (\ref{p1-1}) for any non-zero
potential, even if
Eqs.~(\ref{p1-3}) and (\ref{p1-5}) are satisfied. Therefore, ansatz (\ref{3}%
) always gives only an approximate description of the system. For
the description to be exact, two-particle and all higher-order
correlations in $\Psi _{N}(\mathbf{r}_{1},\ldots ,\mathbf{r}_{N},t)$
must be taken into account~%
\cite{leggett2001,yuv1,gross1962,woo1972,feenberg1974}.

\subsection{Operator method}

T. Wu \cite{wu1961} proposed a method for describing a system of
point bosons, which is based on the ansatz $\hat{\Psi}(\mathbf{r},t)=\hat{a}%
_{0}\psi (\mathbf{r},t)+\hat{\vartheta}(\mathbf{r},t)$  with $\hat{\vartheta%
}(\mathbf{r},t)\ll \hat{a}_{0}\psi (\mathbf{r},t)$. The analysis
\cite{wu1961} resulted in Eqs.~(\ref{p1-3}), (\ref{p1-5}) with
$U(|\mathbf{r}_{j}-\mathbf{r}_{l}|)\rightarrow 2c\delta
(\mathbf{r}_{j}-\mathbf{r}_{l})$ and  $N-1 \rightarrow N$. We are
unable to reproduce Wu's analysis, so we will
make an independent calculation in the simpler case $\hat{\Psi}(\mathbf{r%
},t)=\hat{a}_{0}\psi (\mathbf{r},t)$ with the normalisation $\int_{V}d%
\mathbf{r}|\psi (\mathbf{r},t)|^{2}=1$. Thus, we assume that all $N$ atoms
at any time are in the condensate $\psi (\mathbf{r},t)$, but we do not
change to the c-number. Substituting $\hat{\Psi}(\mathbf{r},t)=\hat{a}%
_{0}\psi (\mathbf{r},t)$ into the Heisenberg equation%
\begin{equation}
i\hbar \frac{\partial \hat{\Psi}(\mathbf{r},t)}{\partial t}=-\frac{\hbar ^{2}%
}{2m}\frac{\partial ^{2}\hat{\Psi}(\mathbf{r},t)}{\partial \mathbf{r}^{2}}%
+\int\limits_{V}d\mathbf{r}^{\prime }U(|\mathbf{r}-\mathbf{r}^{\prime }|)%
\hat{\Psi}^{+}(\mathbf{r}^{\prime },t)\hat{\Psi}(\mathbf{r}^{\prime
},t) \hat{\Psi}(\mathbf{r},t),
\end{equation}%
we obtain%
\begin{equation}
\hat{G}=0,  \label{p1-10b}
\end{equation}%
where%
\begin{eqnarray}
\hat{G} &=&-\hat{a}_{0}i\hbar \frac{\partial \psi (\mathbf{r},t)}{\partial t}%
-\psi (\mathbf{r},t)i\hbar \frac{\partial \hat{a}_{0}}{\partial t}-\hat{a}%
_{0}\frac{\hbar ^{2}}{2m}\frac{\partial ^{2}\psi (\mathbf{r},t)}{\partial
\mathbf{r}^{2}}+  \notag \\
&&+\hat{a}_{0}^{+}\hat{a}_{0}^{2}\psi (\mathbf{r},t)\int\limits_{V}d\mathbf{r%
}^{\prime }U(|\mathbf{r}-\mathbf{r}^{\prime }|)|\psi (\mathbf{r}^{\prime
},t)|^{2}.  \label{p1-10c}
\end{eqnarray}

Consider an $N$-particle state $|N_{0},N_{1},N_{2},\ldots ,N_{\infty
}\rangle $, where $N_{0}=N$ particles are in the one-particle state $\psi
_{0}(\mathbf{r},t)\equiv \psi (\mathbf{r},t)$, whereas the other
one-particle states $\psi _{j}$ from the expansion $\hat{\Psi}(\mathbf{r}%
,t)=\sum_{j=0,1,\ldots ,\infty }\hat{a}_{j}(t)\psi _{j}(\mathbf{r},t)$ are
non-occupied, $N_{j\geq 1}=0$. Denote $|N,0,\ldots ,0\rangle \equiv
|N\rangle $. The expression
\begin{equation}
\hat{G}|N\rangle =0  \label{p1-10d}
\end{equation}%
sets the equation for the condensate $\psi (\mathbf{r},t)$ for the
system in
the $|N\rangle $ state. To find $\hat{G}|N\rangle $, we have to calculate $%
\partial \hat{a}_{0}/\partial t$. If $\hat{\Psi}(\mathbf{r},t)=\hat{a}%
_{0}\psi (\mathbf{r},t)$, the system Hamiltonian takes the form%
\begin{eqnarray}
\hat{H} &=&-\frac{\hbar ^{2}}{2m}\int d\mathbf{r}\hat{\Psi}^{+}(\mathbf{r},t)%
\frac{\partial ^{2}\hat{\Psi}(\mathbf{r},t)}{\partial \mathbf{r}^{2}}+\frac{1%
}{2}\int d\mathbf{r}d\mathbf{r}^{\prime }U(|\mathbf{r}-\mathbf{r}^{\prime }|)%
\hat{\Psi}^{+}(\mathbf{r},t)\hat{\Psi}^{+}(\mathbf{r}^{\prime },t)\hat{\Psi}(%
\mathbf{r}^{\prime },t)\hat{\Psi}(\mathbf{r},t)  \notag \\
&=&E_{k}\hat{a}_{0}^{+}(t)\hat{a}_{0}(t)+E_{p}[\hat{a}_{0}^{+}(t)]^{2}\hat{a}%
_{0}^{2}(t)=(E_{k}-E_{p})\hat{N}_{0}(t)+E_{p}\hat{N}_{0}^{2}(t),
\label{p1-15}
\end{eqnarray}%
where $\hat{N}_{0}(t)=\hat{a}_{0}^{+}(t)\hat{a}_{0}(t)$,
\begin{equation}
E_{k}=-\frac{\hbar ^{2}}{2m}\int d\mathbf{r}\psi ^{\ast }(\mathbf{r},t)\frac{%
\partial ^{2}\psi (\mathbf{r},t)}{\partial \mathbf{r}^{2}},  \label{p1-16}
\end{equation}%
\begin{equation}
E_{p}=\frac{1}{2}\int d\mathbf{r}d\mathbf{r}^{\prime }U(|\mathbf{r}-\mathbf{r%
}^{\prime }|)|\psi (\mathbf{r},t)|^{2}|\psi (\mathbf{r}^{\prime },t)|^{2}.
\label{p1-17}
\end{equation}%
Such a Hamiltonian does not contain terms that could transfer atoms
from the condensate to other states $\psi _{j\geq 1}(\mathbf{r},t)$.
Therefore it is natural to expect that $\frac{\partial
\hat{a}_{0}}{\partial t}=0$.

Let us show that really $\frac{\partial \hat{a}_{0}}{\partial t}=0$.
The derivative $\frac{\partial \hat{a}_{0}}{\partial t}$ cannot be
found from
the Heisenberg equation $i\hbar \frac{\partial \hat{a}_{0}}{\partial t}=[%
\hat{a}_{0},\hat{H}]$. Indeed, from the equations $i\hbar
\frac{\partial
\hat{a}_{j}}{\partial t}=[\hat{a}_{j},\hat{H}]$ ($j=0,1,\ldots ,\infty $), $%
\hat{\Psi}(\mathbf{r},t)=\sum_{j=0,1,\ldots ,\infty }\hat{a}_{j}(t)\psi _{j}(%
\mathbf{r},t)$, and $i\hbar \frac{\partial \hat{\Psi}}{\partial t}=[\hat{\Psi%
},\hat{H}]$, it follows that $\sum_{j=0,1,\ldots ,\infty }\hat{a}_{j}(t)%
\frac{\partial \psi _{j}(\mathbf{r},t)}{\partial t}=0$, i.e. $%
\sum_{j=0,1,\ldots ,\infty }\hat{a}_{j}^{+}(t)\frac{\partial \psi _{j}^{\ast
}(\mathbf{r},t)}{\partial t}=0$. The last two equations must hold for any
state $|N_{0},N_{1},\ldots ,N_{\infty }\rangle $. This means that $\frac{%
\partial \psi _{j}^{\ast }(\mathbf{r},t)}{\partial t}=0$ and $\frac{\partial
\psi _{j}(\mathbf{r},t)}{\partial t}=0$ for all $j=0,1,\ldots ,\infty $,
which contradicts our scheme.

Similarly to the analysis in \cite{wu1961}, let us determine $\frac{\partial
\hat{a}_{0}}{\partial t}$ from the time evolution of the wave function in
the Schr\"{o}dinger representation: $\Psi (\mathbf{r},t)=e^{-i%
\hat{H}t/\hbar }\Psi (\mathbf{r})$~\cite{land3}. Then
\begin{equation}
\Psi (\mathbf{r},t+\delta t)=e^{-i\hat{H}\delta t/\hbar }\Psi (\mathbf{r},t).
\label{p1-12}
\end{equation}%
In the second quantization formalism, the state $|N,0,0,\ldots
,0\rangle $ is \cite{land3}
\begin{equation}
\Psi (t)=(N!)^{-1/2}[\hat{a}_{0}^{+}(t)]^{N}|0\rangle ,  \label{p1-13}
\end{equation}%
where $|0\rangle \equiv |0,0,0,\ldots ,0\rangle $ is the vacuum state:
\begin{equation}
\hat{a}_{j}|0\rangle =0,\quad j=0,1,2,\ldots ,\infty .  \label{vac}
\end{equation}%
Then
\begin{eqnarray}
\Psi (t+\delta t) &\equiv &(N!)^{-1/2}[\hat{a}_{0}^{+}(t+\delta
t)]^{N}|0\rangle =e^{-i\hat{H}\delta t/\hbar }\Psi (t)=(N!)^{-1/2}e^{-i\hat{H%
}\delta t/\hbar }[\hat{a}_{0}^{+}(t)]^{N}|0\rangle  \notag \\
&=&(N!)^{-1/2}\left( 1-i\hat{H}\delta t/\hbar +\frac{1}{2!}(-i\hat{H}\delta
t/\hbar )^{2}+\ldots \right) [\hat{a}_{0}^{+}(t)]^{N}|0\rangle .
\label{p1-14}
\end{eqnarray}%
Note that the energy and the total number of particles are integrals
of motion; therefore, $\hat{H}$ and $\hat{N}$ do not depend on time.
Using the relations
\begin{equation}
\hat{a}_{0}(t)\hat{a}_{0}^{+}(t)=\hat{a}_{0}^{+}(t)\hat{a}_{0}(t)+1,
\label{p1-18}
\end{equation}%
\begin{equation}
\hat{N}_{0}(t)[\hat{a}_{0}^{+}(t)]^{N}=[\hat{a}_{0}^{+}(t)]^{N}(\hat{N}%
_{0}(t)+N),  \label{p1-19}
\end{equation}%
\begin{equation}
\hat{N}_{0}(t)^{2}[\hat{a}_{0}^{+}(t)]^{N}=[\hat{a}_{0}^{+}(t)]^{N}(\hat{N}%
_{0}(t)+N)^{2},  \label{p1-20}
\end{equation}%
and formula (\ref{p1-15}), we obtain
\begin{equation}
(-\hat{H}\delta t/\hbar )^{p}[\hat{a}_{0}^{+}(t)]^{N}=[\hat{a}%
_{0}^{+}(t)]^{N}\hat{f}^{p},  \label{p1-21}
\end{equation}%
\begin{equation}
\hat{f}=(-\delta t/\hbar )\left[ (E_{k}-E_{p})(\hat{N}_{0}(t)+N)+E_{p}(\hat{N%
}_{0}(t)+N)^{2}\right] ,  \label{p1-22}
\end{equation}%
\begin{eqnarray}
&&(N!)^{-1/2}[\hat{a}_{0}^{+}(t+\delta t)]^{N}|0\rangle =(N!)^{-1/2}[\hat{a}%
_{0}^{+}(t)]^{N}e^{i\hat{f}}|0\rangle  \notag \\
&=&(N!)^{-1/2}[\hat{a}_{0}^{+}(t)]^{N}\left( 1+i\hat{f}+\frac{1}{2!}(i\hat{f}%
)^{2}+\ldots \right) |0\rangle .  \label{p1-23}
\end{eqnarray}%
Since
\begin{equation}
i\hat{f}|0\rangle =(-i\delta t/\hbar )\left[ (E_{k}-E_{p})N+E_{p}N^{2}\right]
|0\rangle \equiv ig\delta t|0\rangle ,  \label{p1-24}
\end{equation}%
we have that for an arbitrary $\delta t\geq 0$,%
\begin{equation}
(N!)^{-1/2}[\hat{a}_{0}^{+}(t+\delta t)]^{N}|0\rangle =e^{ig\delta
t}(N!)^{-1/2}[\hat{a}_{0}^{+}(t)]^{N}|0\rangle .  \label{p1-25}
\end{equation}%
To find the correct normalisation, we must replace $e^{ig\delta t}$
by $1$. Eventually,
\begin{equation}
(N!)^{-1/2}[\hat{a}_{0}^{+}(t+\delta t)]^{N}|0\rangle =(N!)^{-1/2}[\hat{a}%
_{0}^{+}(t)]^{N}|0\rangle ,  \label{p1-26}
\end{equation}%
\begin{equation}
\hat{a}_{0}^{+}(t+\delta t)=\hat{a}_{0}^{+}(t),\quad \partial \hat{a}%
_{0}^{+}/\partial t=0\quad \Rightarrow \ \partial \hat{a}_{0}/\partial t=0.
\label{p1-27}
\end{equation}%
The relations (\ref{p1-10c}), $\hat{a}_{0}|N\rangle =\sqrt{N}%
|N-1\rangle $, $\hat{a}_{0}^{+}\hat{a}_{0}|N-1\rangle
=(N-1)|N-1\rangle $, and $\partial \hat{a}_{0}/\partial t=0$ yield
\begin{eqnarray}
&&\hat{G}|N\rangle /\sqrt{N}=\left[ -i\hbar \frac{\partial \psi (\mathbf{r}%
,t)}{\partial t}-\frac{\hbar ^{2}}{2m}\frac{\partial ^{2}\psi (\mathbf{r},t)%
}{\partial \mathbf{r}^{2}}\right. +  \notag \\
&&\left. +(N-1)\psi (\mathbf{r},t)\int\limits_{V}d\mathbf{r}^{\prime }U(|%
\mathbf{\ r}-\mathbf{r}^{\prime }|)|\psi (\mathbf{r}^{\prime },t)|^{2}\right]
|N-1\rangle .  \label{p1-30}
\end{eqnarray}%
For any $\mathbf{r}$ and $t$, the equality $\hat{G}|N\rangle =0$
must hold, which gives the desired equation for the condensate:
\begin{equation}
i\hbar \frac{\partial \psi (\mathbf{r},t)}{\partial t}=-\frac{\hbar ^{2}}{2m}%
\frac{\partial ^{2}\psi (\mathbf{r},t)}{\partial \mathbf{r}^{2}}+(N-1)\psi (%
\mathbf{r},t)\int\limits_{V}d\mathbf{r}^{\prime }U(|\mathbf{r}-\mathbf{r}%
^{\prime }|)|\psi (\mathbf{r}^{\prime },t)|^{2}.  \label{p1-28}
\end{equation}%
This equation coincides with (\ref{p1-7}). So, we again arrive at
equations (\ref{p1-6}) and (\ref{p1-8}).

In work~\cite{wu1961},  different
results were obtained from Eqs.~(\ref{p1-13}) and (\ref{p1-14}) above: namely, $%
\frac{\partial \hat{a}_{0}^{+}}{\partial
t}|_{\hat{\vartheta}\rightarrow 0}\neq 0$, and Eqs.~(\ref{p1-3}),
(\ref{p1-5}) with the replacements $U(|\mathbf{r}_{j}-\mathbf{r}_{l}|)\rightarrow 2c\delta (\mathbf{r}%
_{j}-\mathbf{r}_{l})$ and  $N-1 \rightarrow N $. We are not able to
grasp how the equation for the condensate was derived
in~\cite{wu1961}. The different result
for $\frac{\partial \hat{a}_{0}^{+}}{\partial t}|_{\hat{\vartheta}%
\rightarrow 0}$ may have been obtained  for one or more of the
following reasons:
(i) It was assumed in \cite{wu1961} that $\hat{\Psi}(\mathbf{r},t)=\hat{a}%
_{0}\psi (\mathbf{r},t)+\hat{\vartheta}(\mathbf{r},t)$ instead of
$\hat{\Psi}(\mathbf{r},t) = \hat{a}_{0}\psi(\mathbf{r},t)$.
(ii) Vacuum was defined in work~\cite{wu1961} differently: $\hat{\Psi}(%
\mathbf{r},t)|0\rangle =0$. In this case, formally, $\hat{\vartheta}(\mathbf{%
r},t)|0\rangle =-\hat{a}_{0}\psi (\mathbf{r},t)|0\rangle \neq 0$,
and this equality seems to be applied in \cite[see formulae (2.18),
(2.20), (2.21),
(A10), and (A11)]{wu1961}. However, if $\hat{\vartheta}(\mathbf{r},t)=-\hat{a%
}_{0}\psi (\mathbf{r},t)$, then the smallness condition $\hat{\vartheta}(\mathbf{r%
},t)\ll \hat{a}_{0}\psi (\mathbf{r},t)$, which was used in
\cite{wu1961} to develop the perturbation theory, is violated.
Moreover, in the second quantization formalism~\cite{land3}, vacuum is the state $%
|0,0,\ldots ,0\rangle $ corresponding to the occupation numbers $N_{j}=0$, $%
j=0,1,\ldots ,\infty $, from which condition~(\ref{vac}) follows. It
is not difficult to show that the equality
$\hat{\Psi}(\mathbf{r},t)|0\rangle =0$ is possible only if
condition~(\ref{vac}) holds. (iii)~In work~\cite{wu1961}, the
expression for $\Psi (t+\delta t)$ was not written in the form that
allows one to extract the terms which give a zero contribution when
acting on the vacuum. (iv) Apparently, Wu \cite{wu1961} considered
 only the case of $N\gg 1$, then $N-1 \approx N $.

In any event, Wu proposed the right idea that the equation for a
nonuniform condensate can be obtained without going to the c-number.
Moreover, Wu derived an equation which is equivalent to the GP
equation, simultaneously with Pitaevskii~\cite{pit1961} and
Gross~\cite{gross1961} and using the more precise ansatz
$\hat{\Psi}(\mathbf{r},t) = \hat{a}_{0}\psi(\mathbf{r},t) +
\hat{\vartheta}(\mathbf{r},t)$.

Note that the replacement of an operator by a c-number ($\hat{\psi}(x,t)=%
\hat{a}_{0}\Psi (\mathbf{r},t)/\sqrt{N}\rightarrow \Psi (x,t)$)
creates the uncertainty $\pm 1$ for the number of particles and thus
violates the conservation law for this parameter. To solve this
difficulty, N.~Bogoliubov proposed the method of
quasi-averages~\cite{bogquasi}; this method is actually reduced to
the mechanism of spontaneous symmetry breaking (SSB), which removes
statistical degeneracy. However, this is a purely formal trick. In
nature, the SSB occurs in a different way, under a phase transition,
which is usually initiated by the formation of new phase nuclei.
It is important that the application of the operator ansatz $\hat{\Psi}%
(\mathbf{r},t)=\hat{a}_{0}\Psi (\mathbf{r},t)/\sqrt{N}$
automatically eliminates the difficulty with the conservation law
for $N$.

Thus,  using the operator $\hat{\Psi}(\mathbf{r},t)=\hat{a}%
_{0}\Psi (\mathbf{r},t)/\sqrt{N}$ instead of the c-number $\hat{\Psi}(%
\mathbf{r},t)=\Psi (\mathbf{r},t)$ results in Eq.~(\ref{p1-8}),
which contains the additional factor $\left( 1-1/N\right) $ in
comparison with the ordinary Gross-Pitaevskii equation. This
multiplier also appears in
the approach, which rests on the ansatz $\Psi _{N}(\mathbf{r}_{1},\ldots ,\mathbf{r}%
_{N},t)=\prod_{j=1}^{N}\psi (\mathbf{r}_{j},t)$. The approximation $\hat{\Psi%
}(\mathbf{r},t)=\Psi (\mathbf{r},t)$ requires that $N\gg 1$, but the
ans\"{a}tze $\hat{\Psi}(\mathbf{r},t)=\hat{a}_{0}\Psi
(\mathbf{r},t)/\sqrt{N}$
and $\Psi _{N}(\mathbf{r}_{1},\ldots ,\mathbf{r}_{N},t)=\prod_{j=1}^{N}\psi (%
\mathbf{r}_{j},t)$ are valid for any $N\geq 2$. Therefore, using such
ans\"{a}tze instead of the c-number $\hat{\Psi}(\mathbf{r},t)=\Psi (%
\mathbf{r},t)$ makes it possible to extend the domain of
applicability of the Gross-Pitaevskii equation to small $N$.

\section{Ground state of condensate: equations and numerical method}

In order to be able to compare the solutions with exact ones, let us
consider a 1D system. We will use zero BCs because in this case the
particle density $\rho (x)$ depends on the coordinate (under
periodic BCs, $\rho (x)=\mathrm{const}$). So, let us consider $N$
spinless Bose particles, which occupy the segment $[0,L]$, under
zero BCs ($\Psi (0,t)=\Psi (L,t)=0$). We assume that the interaction
is repulsive, and the wave function of the condensate considerably
changes on scales much larger
than the atomic size. Therefore, we use the GP equation (\ref{2}) and the GP$%
_{N}$ equation (\ref{p1-8}) instead of Eqs.~(\ref{1}) and
(\ref{p1-6}). We
seek stationary solutions:%
\begin{equation}
\Psi (x,t)=e^{\epsilon t/i\hbar }\Phi (x).  \label{8}
\end{equation}%
Then the GP and GP$_{N}$ equations (\ref{2}) and (\ref{p1-8}) take the form%
\begin{equation}
\epsilon \Phi (x)=-\frac{\hbar ^{2}}{2m}\frac{\partial ^{2}\Phi }{\partial
x^{2}}+2c|\Phi |^{2}\Phi ,  \label{9-1}
\end{equation}%
\begin{equation}
\epsilon \Phi (x)=-\frac{\hbar ^{2}}{2m}\frac{\partial ^{2}\Phi
}{\partial x^{2}}+\left( 1-\frac{1}{N}\right) 2c|\Phi |^{2}\Phi,
\label{9-2}
\end{equation}%
respectively, with the boundary conditions%
\begin{equation}
\Phi (x=0)=\Phi (x=L)=0.  \label{bc2}
\end{equation}

Let the condensate $\Psi (x,t)$ contain $N_{c}$ atoms, then%
\begin{equation}
\int\limits_{0}^{L}dx|\Phi (x)|^{2}=N_{c}.  \label{norm}
\end{equation}%
Since the condensate approximations $\hat{\Psi}(x,t)=\Psi (x,t)$, (\ref{3}),
and (\ref{5}) mean that $N_{c}=N$, we put $N_{c}=N$.

To satisfy BCs~(\ref{bc2}), we will seek each solution of the GP
(GP$_{N}$) equation as an expansion in the complete orthonormal set
of sines,
\begin{equation}
\Phi (x)=\sum\limits_{j=1,2,\ldots ,\infty }b_{j}\sqrt{2/L}\cdot \sin
(k_{j}x),\quad k_{j}=\pi j/L.  \label{10}
\end{equation}%
One can  see that there are \textquotedblleft elementary $j_{0}$%
-series\textquotedblright\
\begin{equation}
\Phi _{j_{0}}(x)=\sum\limits_{j=j_{0},3j_{0},5j_{0},\ldots }b_{j}\sqrt{2/L}%
\cdot \sin (k_{j}x),  \label{11}
\end{equation}%
for which $j_{0}$ can be equal to $1,2,3,\ldots ,\infty $. If the
$j_{0}$-series (\ref{11}) is substituted into Eq.~(\ref{9-1}) or
(\ref{9-2}), then the
right-hand and left-hand sides will only contain terms with the structure of the same $%
j_{0}$-series (\ref{11}) (in this case, each product of three sine
functions should be represented as a sum of sines).

In the absence of interaction ($c=0$), Eqs.~(\ref{9-1}) and (\ref{9-2}) with
BCs~(\ref{bc2}) have the solutions $\Phi (x)=\sqrt{2N/L}\cdot \sin (k_{j_{0}}x)$%
, $j_{0}=1,2,3,\ldots $, which form a complete set of functions. If
$c\neq 0$, these solutions transform into the series~(\ref{11}) with
$j_{0}=1,2,3,\ldots $ and $b_{j\neq j_{0}}\neq 0$. In this paper and
\cite{gp2}, we analyse only solutions in the form of the
$j_{0}$-series (\ref{11}). For each value of $j_{0}$ we have found
one and only one solution~$\Phi (x)$. Below we will see that the
$j_{0}$-series corresponds to the particle density profile $\rho
(x)$ with $j_{0}$ domains. According to the results of
work~\cite{carr2000r}, such solutions include all solutions of the
GP equation with zero BCs. The ground state of the condensate
corresponds to a single-domain solution ($j_{0}=1$).

Let us substitute expansion (\ref{11}) into Eq.~(\ref{9-2}) and
express each product of three sines as a sum of sines. Then we
obtain the equation
\begin{eqnarray}
&&\sum\limits_{j}\left( \epsilon -\frac{\hbar ^{2}k_{j}^{2}}{2m}\right)
b_{j}\sin (k_{j}x)=\left( 1-\frac{1}{N}\right) \frac{c}{L}%
\sum_{j_{1}j_{2}j_{3}}b_{j_{1}}^{\ast }b_{j_{2}}b_{j_{3}}\left[ \sin
(k_{j_{3}-j_{1}+j_{2}}x)+\right.  \notag \\
&&+\left. \sin (k_{j_{3}+j_{1}-j_{2}}x)-\sin (k_{j_{3}-j_{1}-j_{2}}x)-\sin
(k_{j_{3}+j_{1}+j_{2}}x)\right] ,  \label{12}
\end{eqnarray}%
where $j,j_{1},j_{2},j_{3}$ run over the values
$j_{0},3j_{0},5j_{0},\ldots ,\infty $. Denote by $j$ each index of
the form $j_{3}\pm j_{1}\pm j_{2}$ on the right-hand side of
Eq.~(\ref{12}) and pass from summation over $j_{3}$ to summation
over $j=\pm j_{0},\pm 3j_{0},\pm 5j_{0},\ldots $. The descriptions
in terms of the subscripts $j_{3}$ and $j$ are equivalent. Further,
for all $j<0$ we take into account the property $\sin (k_{j})=-\sin
(k_{-j})$, make the substitution $j=-\tilde{j}$, and omit the tilde.
As a result, Eq.~(\ref{12}) takes the form
\begin{eqnarray}
&&\sum\limits_{j}\left( \epsilon -\frac{\hbar ^{2}k_{j}^{2}}{2m}\right)
b_{j}\sin (k_{j}x)=\left( 1-\frac{1}{N}\right) \frac{c}{L}%
\sum_{j_{1},j_{2},j}b_{j_{1}}^{\ast }b_{j_{2}}\sin (k_{j}x)\left[ \theta
(j+j_{1}-j_{2})b_{j+j_{1}-j_{2}}-\right.  \notag \\
&&-\theta (-j+j_{1}-j_{2})b_{-j+j_{1}-j_{2}}+\theta
(j-j_{1}+j_{2})b_{j-j_{1}+j_{2}}-\theta (-j-j_{1}+j_{2})b_{-j-j_{1}+j_{2}}-
\notag \\
&&-\left. b_{j+j_{1}+j_{2}}+\theta (-j+j_{1}+j_{2})b_{-j+j_{1}+j_{2}}-\theta
(j-j_{1}-j_{2})b_{j-j_{1}-j_{2}}\right] ,  \label{12b}
\end{eqnarray}%
where $j,j_{1},j_{2}=j_{0},3j_{0},5j_{0},\ldots ,\infty $, and $\theta (p)$
is the discrete Heaviside function: $\theta (p)=1$ for $%
p\geq 0$, and $\theta (p)=0$ for $p<0$.

Then we collect the coefficients of the independent functions $\sin
(k_{j}x)$ and denote $b_{j}=\sqrt{N}f_{j}$, $N/L=\bar{\rho}$, $\gamma =\frac{2mc}{%
\hbar ^{2}\bar{\rho}}$, and $\epsilon =2\bar{\rho}\left( 1-\frac{1}{N}\right) c\cdot \tilde{%
\epsilon}$. As a result, we arrive at the following nonlinear system
of equations for the unknown $\tilde{\epsilon}$ and the coefficients
$f_{j}$:
\begin{eqnarray}
&&\left( \frac{\pi ^{2}j^{2}}{(1-N^{-1})\gamma N^{2}}-2\tilde{\epsilon}%
\right) f_{j}+\sum_{j_{1}j_{2}}f_{j_{1}}^{\ast }f_{j_{2}}\left[ \theta
(j+j_{1}-j_{2})f_{j+j_{1}-j_{2}}-\right.  \notag \\
&-&\theta (-j+j_{1}-j_{2})f_{-j+j_{1}-j_{2}}+\theta
(j-j_{1}+j_{2})f_{j-j_{1}+j_{2}}-\theta (-j-j_{1}+j_{2})f_{-j-j_{1}+j_{2}}-
\notag \\
&-&\left. f_{j+j_{1}+j_{2}}+\theta (-j+j_{1}+j_{2})f_{-j+j_{1}+j_{2}}-\theta
(j-j_{1}-j_{2})f_{j-j_{1}-j_{2}}\right] =0,  \label{13}
\end{eqnarray}%
where $j,j_{1},j_{2}$ run over the values
$j_{0},3j_{0},5j_{0},\ldots ,\infty $. System (\ref{13}) has to be
supplemented by the normalisation condition following from
(\ref{norm}),
\begin{equation}
\sum\limits_{j=j_{0},3j_{0},5j_{0},\ldots ,\infty }|f_{j}|^{2}=1.
\label{norm2}
\end{equation}

To solve Eqs.~(\ref{13}) and (\ref{norm2}), it is convenient to set $%
j=j_{0}(2l-1)$, $f_{j_{0}(2l-1)}\equiv g_{l}$ and go to the
numbering by the index $l$. Since $j=j_{0},3j_{0},5j_{0},\ldots ,\infty $, we have $%
l=1,2,3,\ldots ,\infty $. Because all solutions $\Phi (x)$ of
Eqs.~(\ref{9-1}) and (\ref{9-2}) can be written in the real-valued
form (see Appendix), we
set $f_{j}^{\ast }=f_{j}$ and $g_{l}^{\ast }=g_{l}$. In fine, Eqs.~(\ref%
{13}) and (\ref{norm2}) take the form
\begin{eqnarray}
&&\left( \frac{\pi ^{2}j_{0}^{2}(2l-1)^{2}}{(1-1/N)\gamma N^{2}}-2\tilde{%
\epsilon}\right) g_{l}+\sum_{l_{1},l_{2}=1,2,\ldots ,\infty
}g_{l_{1}}g_{l_{2}}\left[ \theta
(l+l_{1}-l_{2}-1)g_{l+l_{1}-l_{2}}-\right.
\notag \\
&-&\theta (-l+l_{1}-l_{2})g_{-l+l_{1}-l_{2}+1}+\theta
(l-l_{1}+l_{2}-1)g_{l-l_{1}+l_{2}}-\theta
(-l-l_{1}+l_{2})g_{-l-l_{1}+l_{2}+1}-  \notag \\
&-&\left. g_{l+l_{1}+l_{2}-1}+\theta
(-l+l_{1}+l_{2}-1)g_{-l+l_{1}+l_{2}}-\theta
(l-l_{1}-l_{2})g_{l-l_{1}-l_{2}+1}\right] =0,\quad l=1,2,\ldots ,\infty ,
\label{14}
\end{eqnarray}%
\begin{equation}
\sum\limits_{l=1,2,\ldots ,\infty }g_{l}^{2}=1.  \label{norm3}
\end{equation}%
The system of equations (\ref{14}) is obtained for the GP$_{N}$ equation (%
\ref{9-2}). This system also corresponds to the GP equation
(\ref{9-1}) if we make
the substitution $\left( 1-1/N\right) \rightarrow 1$ in Eq.~(\ref{14}%
) and in the formula $\epsilon =2\bar{\rho}\left( 1-1/N\right) c \tilde{\epsilon%
}$.  In what follows we will solve the system of nonlinear equations
(\ref{14}), (\ref{norm3}) numerically using the Newton method. This
method is well known; for a system of two equations, it is
explained, e.g., in \cite{mt2015}.

The accuracy of the GP and GP$_{N}$ approaches will be verified by
comparing the calculated system energy with the exact energy found
by the Bethe
ansatz. From the second quantization approach and the approximation $\hat{%
\Psi}(x,t)=\Psi (x,t)=e^{\epsilon t/i\hbar }\Phi (x)$, it follows \cite%
{bog1947} that each GP solution $\Phi (x)$ corresponds to the system
energy
\begin{equation}
E_{GP}=\int\limits_{0}^{L}dx\left\{ -\frac{\hbar ^{2}}{2m}\Phi ^{\ast }(x)%
\frac{\partial ^{2}}{\partial x^{2}}\Phi (x)+qc|\Phi (x)|^{4}\right\}
\label{5-01}
\end{equation}%
with $q=1$. Formula (\ref{5-01}) describes the stationary state of
condensate for $N\gg 1$. By using the exact quantum mechanical
formula
\begin{equation}
E=\int\limits_{0}^{L}dx_{1}\ldots dx_{N}\Psi _{N}^{\ast }(x_{1},\ldots
,x_{N})\left[ -\frac{\hbar ^{2}}{2m}\sum\limits_{j=1}^{N}\frac{\partial ^{2}%
}{\partial x_{j}^{2}}+\sum\limits_{j<l}U(|x_{j}-x_{l}|)\right] \Psi
_{N}(x_{1},\ldots ,x_{N})
\end{equation}%
with $U(|\mathbf{r}_{j}-\mathbf{r}_{l}|)= 2c\delta (\mathbf{r}%
_{j}-\mathbf{r}_{l})$ and the condensate ansatz (\ref{3}) with $\psi
(x,t)=e^{\epsilon t/i\hbar }e^{i\kappa(t)/\hbar}\Phi (x)/\sqrt{N}$,
we obtain formula (\ref{5-01}) with $q=1-1/N$.
In the operator approach with $\hat{\Psi}(\mathbf{r},t)=\hat{a}_{0}\Psi (%
\mathbf{r},t)/\sqrt{N}$, the formulae $E=\langle 0,\ldots ,0,N|\hat{H}%
|N,0,\ldots ,0\rangle $, (\ref{p1-15})--(\ref{p1-17}), and $\psi (\mathbf{r}%
,t)=\Psi (\mathbf{r},t)/\sqrt{N}=e^{\epsilon t/i\hbar }\Phi (x)/\sqrt{N}$
again bring about (\ref{5-01}) with $q=1-1/N$. That is, formula (\ref%
{5-01}) with $q=1-1/N$ gives the energy of the stationary condensate
state $\hat{\Psi}(x,t)=\hat{a}_{0}e^{\epsilon t/i\hbar }\Phi
(x)/\sqrt{N}$ (or (\ref{3})) for any $N\geq 2$. This is the energy
in the GP$_{N}$ approach.

Substituting the function $\Phi (x)$~(\ref{11}) with $b_{j}=\sqrt{N}%
f_{j}=\sqrt{N}f_{j_{0}(2l-1)}=\sqrt{N}g_{l}$ into Eq.~(\ref{5-01}),
after some algebra we get
\begin{eqnarray}
E_{GP} &=&\frac{\hbar ^{2}\bar{\rho}^{2}}{2m}\frac{\pi ^{2}j_{0}^{2}}{N}%
\sum\limits_{l=1}^{l_{m}-1}(2l-1)^{2}g_{l}^{2}+\frac{qc \bar{\rho} N}{2}%
\sum\limits_{l_{1}l_{2}l_{3}=1}^{l_{m}-1}g_{l_{1}}g_{l_{2}}g_{l_{3}}\left[
3\theta (l_{1}+l_{2}-l_{3}-1)g_{l_{1}+l_{2}-l_{3}}-\right.  \notag \\
&-&\left. 3\theta (l_{1}-l_{2}-l_{3})g_{l_{1}-l_{2}-l_{3}+1}-\theta
(l_{1}+l_{2}+l_{3}-2)g_{l_{1}+l_{2}+l_{3}-1}\right] .  \label{5-0}
\end{eqnarray}%
Here $q=1$ for the GP approach, and $q=1-1/N$ for the GP$_{N}$
approach.

In the exact approach, the wave functions of a 1D system of point
bosons are given by the Bethe ansatz \cite{gaudin1971,bethe,ll1963}
(see also reviews \cite{gaudinm,syrwid2021}). Under periodic BCs,
the wave function of the 1D system of $N$ point spinless bosons for
the region $x_{1}\leq x_{2}\leq \ldots \leq x_{N}$ is given by the
Bethe ansatz \cite{ll1963}
\begin{equation}
\psi _{\{k\}}(x_{1},\ldots
,x_{N})=\sum\limits_{P}a(P)e^{i\sum\limits_{l=1}^{N}k_{P_{l}}x_{l}},
\label{5-p}
\end{equation}%
where $k_{P_{l}}$ is selected from the set $(k_{1},\ldots ,k_{N})$,
and $P$ means all possible permutations of $k_{l}$. Under zero BCs,
the wave function of the system is a
superposition of a set of counter-propagating waves \cite{gaudin1971,gaudinm}%
,
\begin{equation}
\Psi _{\{|k|\}}(x_{1},\ldots ,x_{N})=\sum\limits_{\varepsilon _{1},\ldots
,\varepsilon _{N}=\pm 1}C(\varepsilon _{1},\ldots ,\varepsilon _{N})\psi
_{\{k\}}(x_{1},\ldots ,x_{N}),  \label{5-1}
\end{equation}%
where $\psi _{\{k\}}$ is defined by formula (\ref{5-p}) with $%
k_{j}=\varepsilon _{j}|k_{j}|$. The formulae for $C(\varepsilon
_{1},\ldots
,\varepsilon _{N})$ and $a(P)$ are written out in \cite%
{gaudin1971,gaudinm,syrwid2021}. The energy of the system of point
bosons
\begin{equation}
E_{\mathrm{Bethe}}=k_{1}^{2}+k_{2}^{2}+\ldots +k_{N}^{2}.  \label{5-2}
\end{equation}

Under zero BCs, the numbers $|k_{j}|$ satisfy the system of Gaudin's equations~%
\cite{gaudin1971,gaudinm},
\begin{equation}
L|k_{p}|=\pi n_{p}+\sum\limits_{j=1}^{N}\left( \arctan {\frac{c}{%
|k_{p}|-|k_{j}|}}+\arctan {\frac{c}{|k_{p}|+|k_{j}|}}\right) |_{j\neq
p},\quad p=1,\ldots ,N,  \label{5-3}
\end{equation}%
where quantum numbers $n_{p}$ are integers, and $n_{p}\geq 1$. The
ground state corresponds to $n_{p}=1$ for all $p=1,2,\ldots ,N$ ($
n_{p\leq N}=1$ for short). The system of equations (\ref{5-3}) has a
unique real solution $\{|k_{p}|\}\equiv (|k_{1}|,|k_{2}|,\ldots
,|k_{N}|)$ for each set $\{n_{p}\}$~\cite{mtjpa2017}. The positivity
of all $|k_{p}|$ has not been proven in \cite{mtjpa2017}, but it can
be corroborated by the direct numerical solution of the system
(\ref{5-3}).

\begin{figure}
\includegraphics[width=0.47\textwidth]{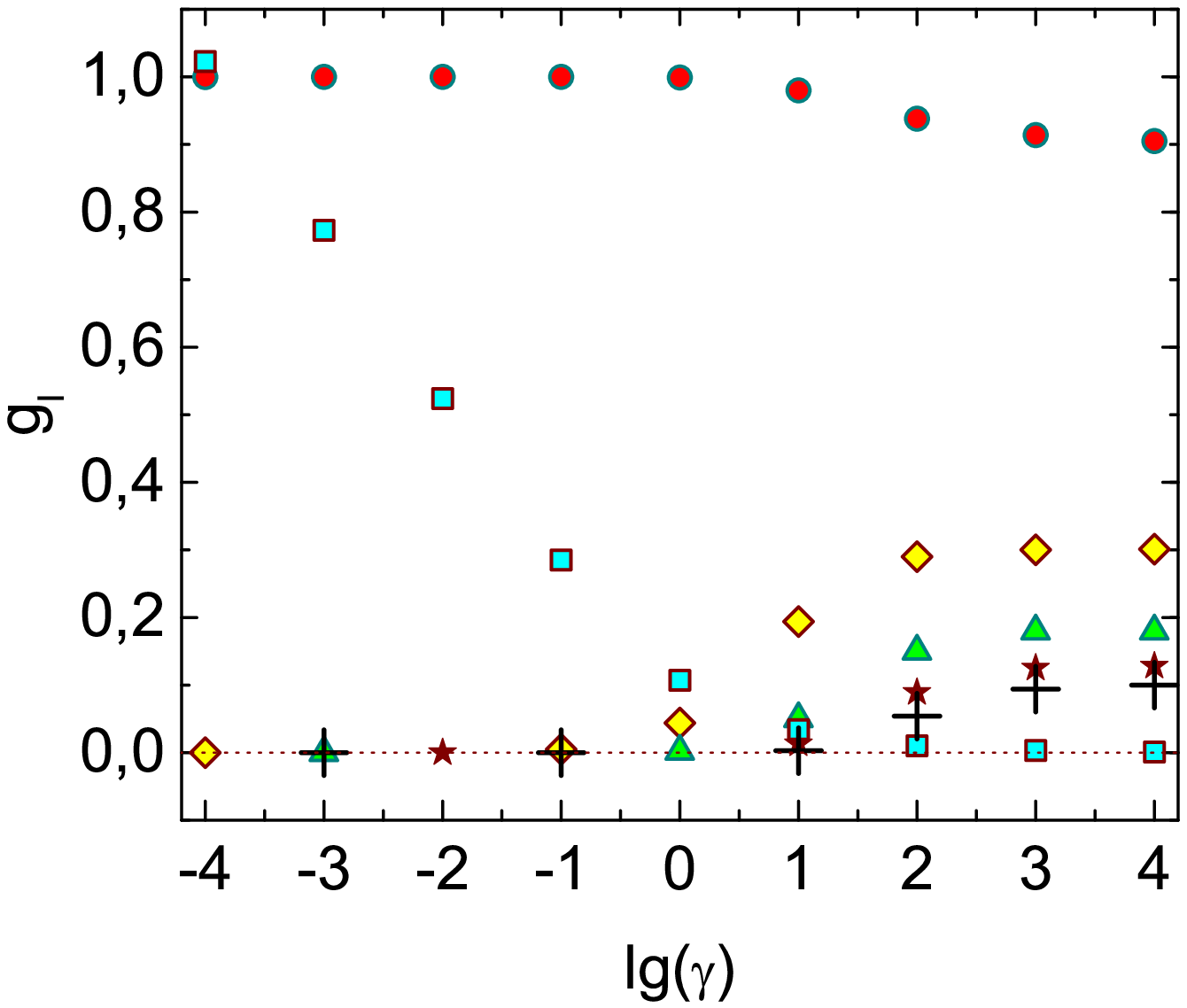}
\hfill
\includegraphics[width=0.47\textwidth]{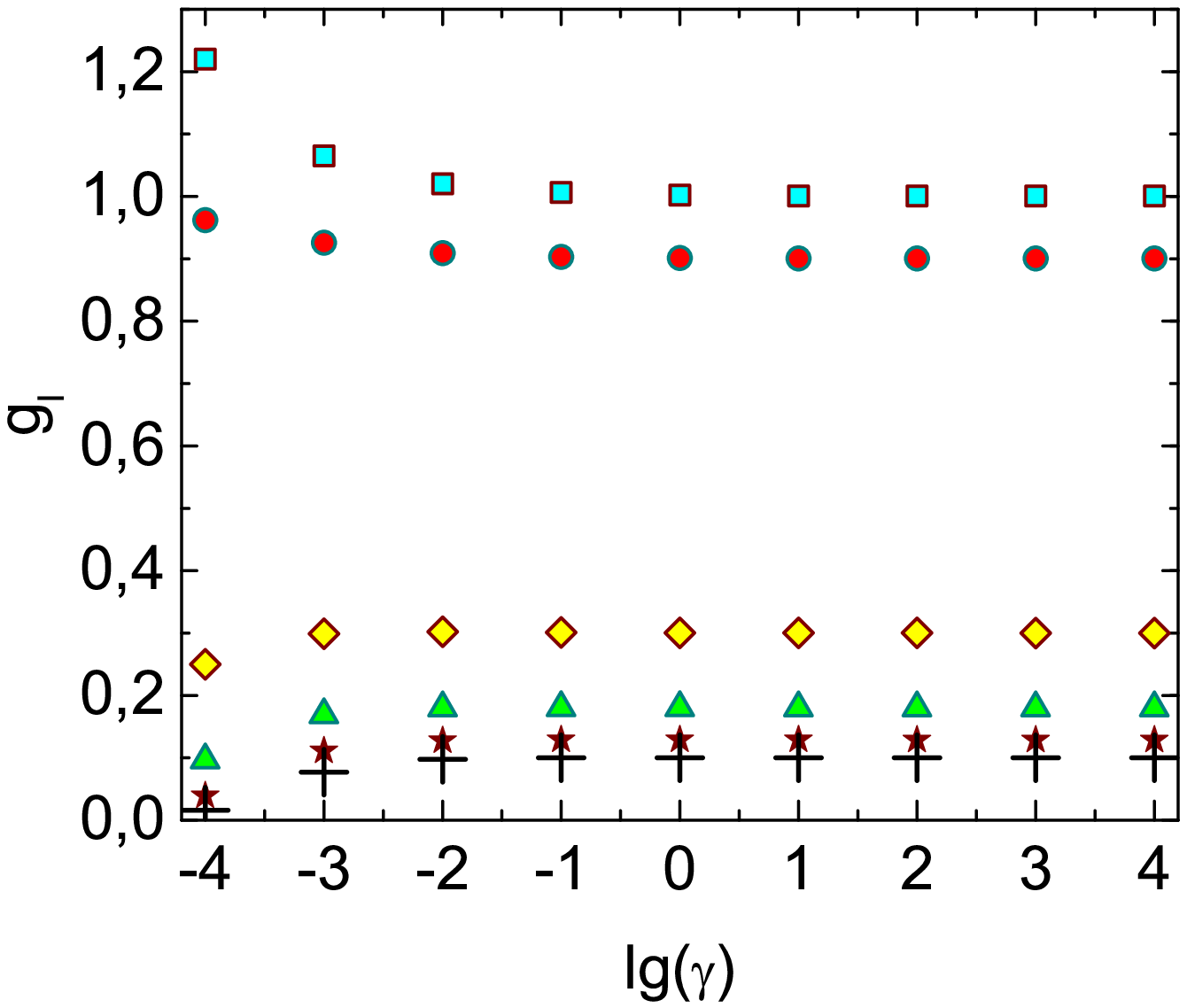}
\\
\parbox[t]{0.47\textwidth}{
\caption{ [Color online] Coefficients $g_{1}$ (circles), $g_{2}$
(diamonds), $g_{3}$  (triangles), $g_{4}$ (asterisks), $g_{5}$
(crosses), and $g_{l_{m}}\equiv \tilde{\epsilon}$ (squares) obtained
within the GP approach for $N=2$, $\bar{\rho}=1$, and various
$\gamma $.  The dotted line marks the zero level $g_{l}=0$.  Shown
are the values of $g_{l}$ for $l=1, 2, 3, 4, 5$ and the values of
$0.25\lg{(g_{l})}$ (instead of $g_{l}$) for $l=l_{m}$. Here and
below, we denote $\log_{10}{(f)}=\lg{f}$.
 \label{fig1gsgN2}} } \hfill
\parbox[t]{0.47\textwidth}{
\caption{ [Color online] Coefficients $g_{1}$ (circles), $g_{2}$
(diamonds), $g_{3}$  (triangles), $g_{4}$ (asterisks), $g_{5}$
(crosses), and $g_{l_{m}}\equiv \tilde{\epsilon}$ (squares)
calculated within the GP approach  for $N=1000$, $\bar{\rho}=1$, and
different $\gamma $.
 \label{fig2gsgN1000}} }
\end{figure}

Below we find the set of numbers $|k_{j}|$ by numerically solving,
using the Newton method, the system of equations (\ref{5-3}) with
$n_{p\leq N}=1$ and various $N$, $L$,  $\gamma $. As a result, we
obtain the exact energy $E_{\mathrm{Bethe}}$ (\ref{5-2}), which
makes it possible to compare it with $E_{\mathrm{GP}}$ and
$E_{\mathrm{GP_{N}}}$ (\ref{5-0}) (in so doing, we must set $\hbar
=2m=1$ in (\ref{5-0}) because formulae (\ref{5-p})--(\ref{5-3}) were
obtained just for this normalisation).

\section{Ground state solutions}

The ground state of the system is described by the function $\Phi
_{j_{0}}(x) $ (\ref{11}) with $j_{0}=1$. In this section we analyse
this solution for different values of $N$ and the coupling constant
$\gamma =c/\bar{\rho}$.

First of all, the wave function of condensate $\Phi _{1}(x)$ (\ref{11}) has
to be determined. Knowing $\Phi _{1}(x)$, we can find the ground state
energy $E_{0}$, the {particle density} profile $\rho (x)$, and other
quantities. We are interested in $E_{0}$ and $\rho (x)$. To find $\Phi
_{1}(x)$ (\ref{11}), it is necessary to solve the system of equations (\ref%
{14}) and (\ref{norm3}) for $j_{0}=1$. We solved it numerically with
the help of the Newton method, putting $l,l_{1},l_{2}=1,2,\ldots
,l_{m}-1$ and denoting $\tilde{\epsilon}\equiv g_{l_{m}}$ (as a
result, we obtained $l_{m}$
equations for $l_{m}$ unknowns $g_{1},g_{2},\ldots ,g_{l_{m}}$). For $j_{0}=1$%
, we found a unique solution of Eqs.~(\ref{14}) and (\ref{norm3})
for each of the considered sets $(N,\bar{\rho},\gamma )$.

\subsection{Coefficients $g_{l}$}

The coefficients $f_{j}=\frac{2\sqrt{2}}{\pi j}$ ($g_{l}=\frac{2\sqrt{2}}{%
\pi (2l-1)}$) and the energy $\tilde{\epsilon}=1$ were used as seed
values for the Newton method. They correspond to the zero
approximation solution
for the Bose gas with weak point interaction ($N^{-2}\ll \gamma \ll 1$)~\cite%
{mtmethodbog}. For the seed $f_{j}$, formula~(\ref{11}) gives $\Phi
_{1}(x)=\sqrt{\bar{\rho}}$ inside the system and $\Phi _{1}(x)=0$ at
the boundaries.

The solutions $g_{l}$ obtained within the GP approach for $N=2$ and
$N=1000$ are shown in Figs.~\ref{fig1gsgN2} and \ref{fig2gsgN1000},
respectively. The values of $g_{l}$ obtained in the GP$_{N}$
approach are close
by magnitude and therefore not shown. At $\gamma \gg 1$ the coefficients $%
g_{l}$ are almost independent of $\gamma $ and close to the seed values $%
g_{l<l_{m}}=\frac{2\sqrt{2}}{\pi (2l-1)}$, $g_{l_{m}}=1$. At $\gamma \ll 1$,
the coefficient $g_{1}$ is close to unity for all $N$. The coefficients $%
g_{1<l<l_{m}}$ decrease as $\gamma $ decreases. At $\gamma \ll 1$
they are smaller for smaller $N$ and differ strongly from the seed values $%
g_{1<l<l_{m}}$. Recall that $g_{l_{m}}$ denotes the energy $\tilde{\epsilon}$%
. The latter decreases with increasing $\gamma $ and becomes close to $%
\tilde{\epsilon}=1$ for all $N$ at $\gamma \gg 1$. At $\gamma \ll 1$
the values of $\tilde{\epsilon}$ increase rapidly as $N$ decreases.

\begin{figure}
\begin{center}
\includegraphics[width=.47\textwidth]{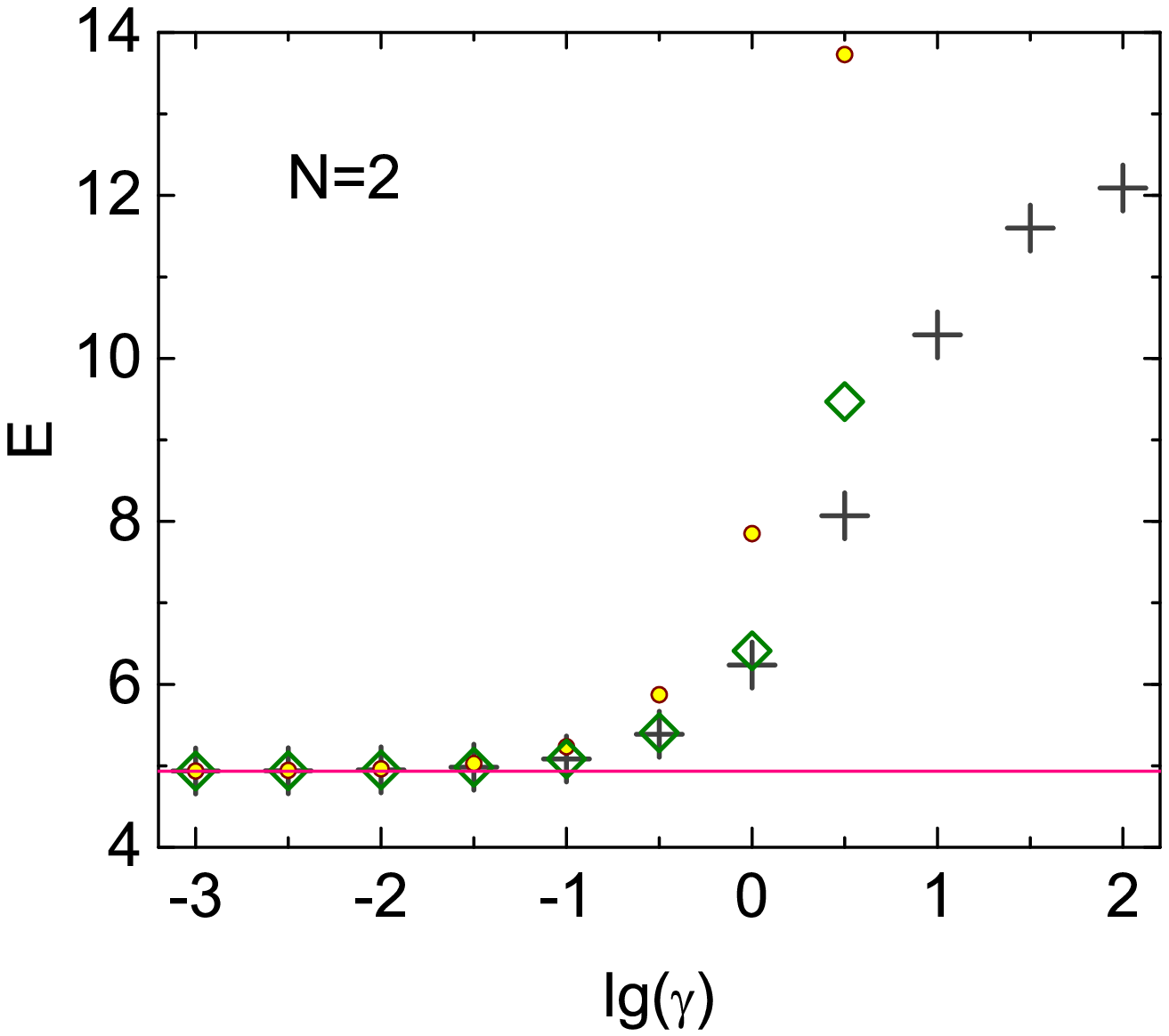}
\hspace{.04\textwidth}
\includegraphics[width=.47\textwidth]{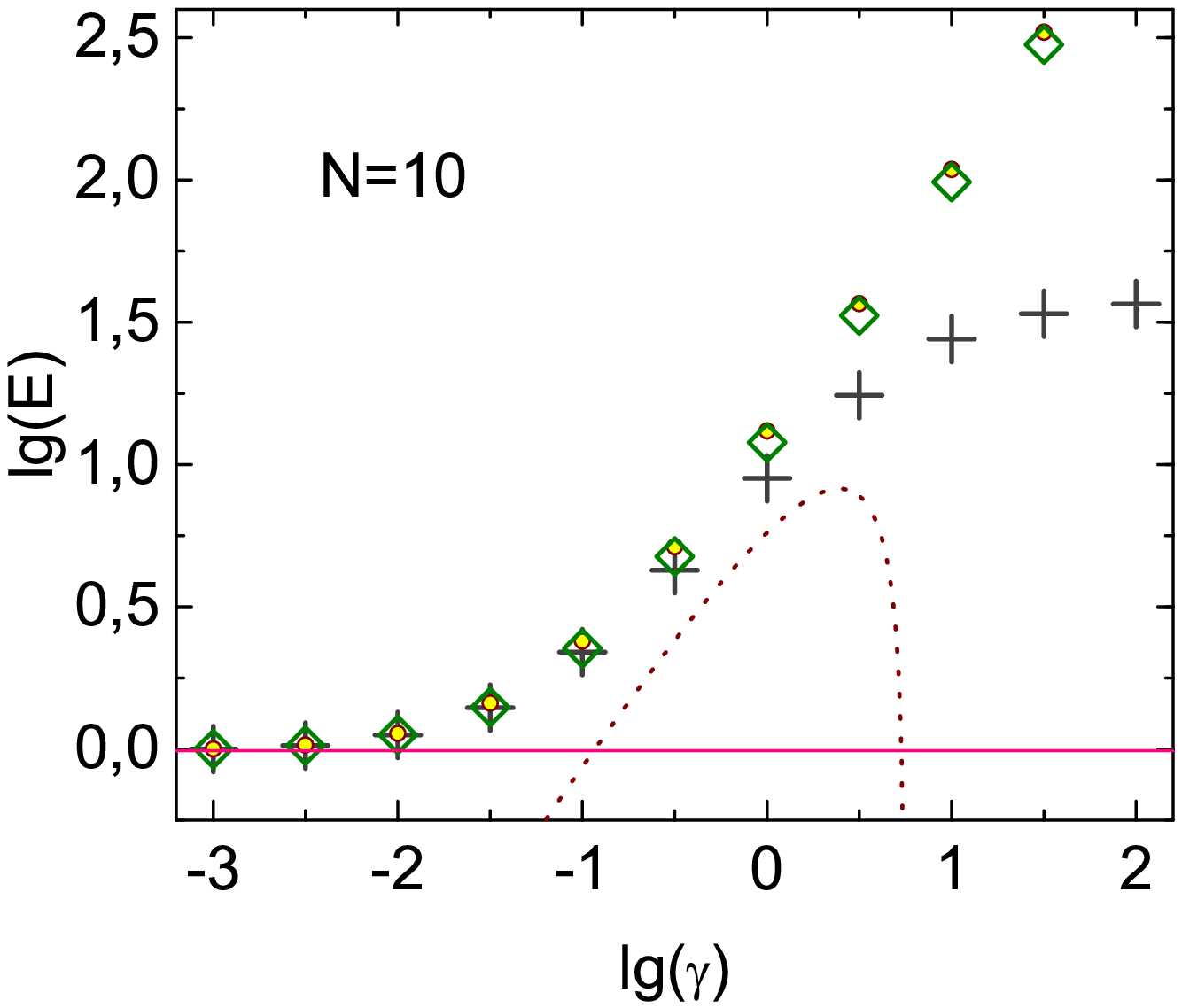}\\
\includegraphics[width=.47\textwidth]{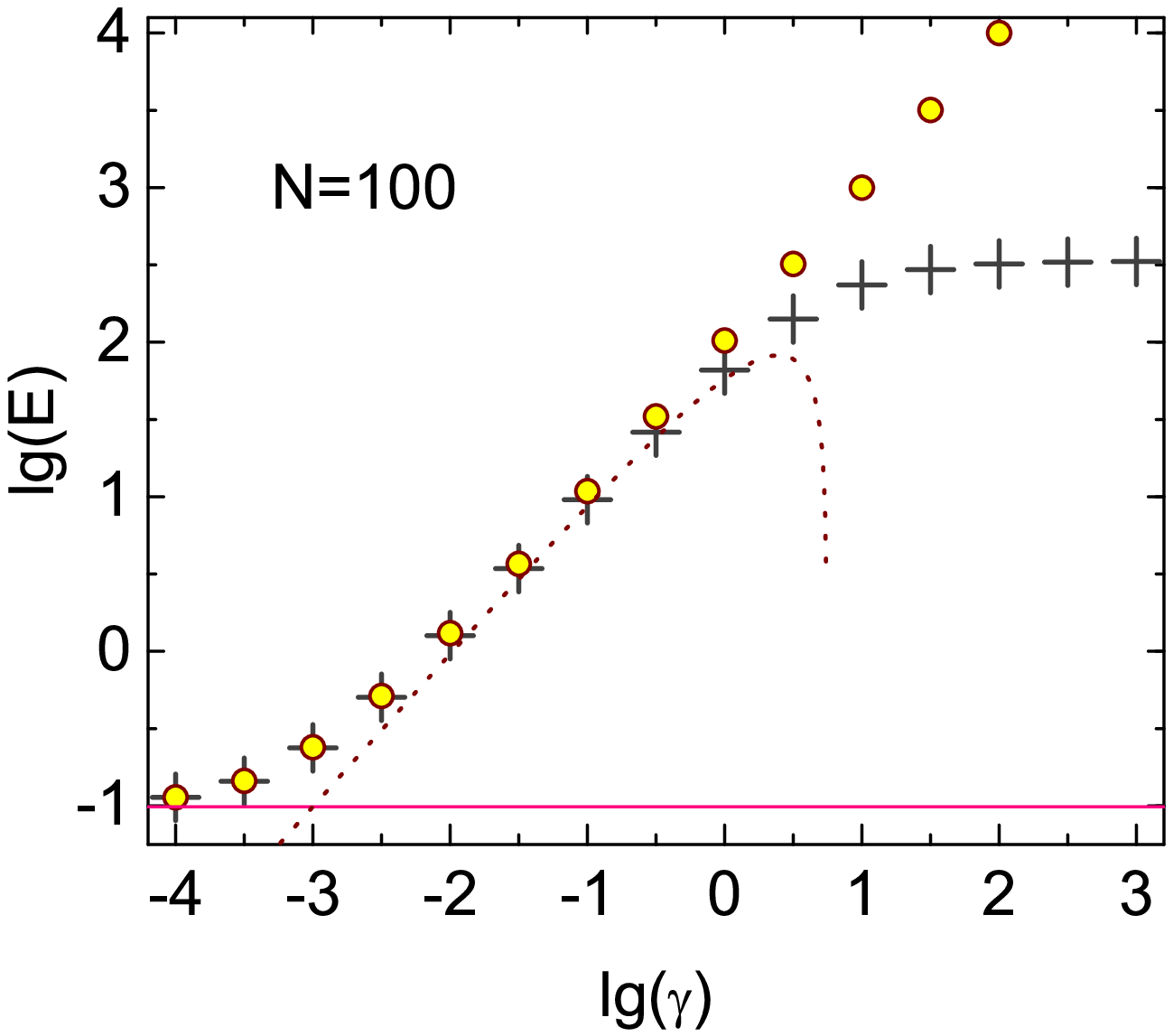}
\hspace{.04\textwidth}
\includegraphics[width=.47\textwidth]{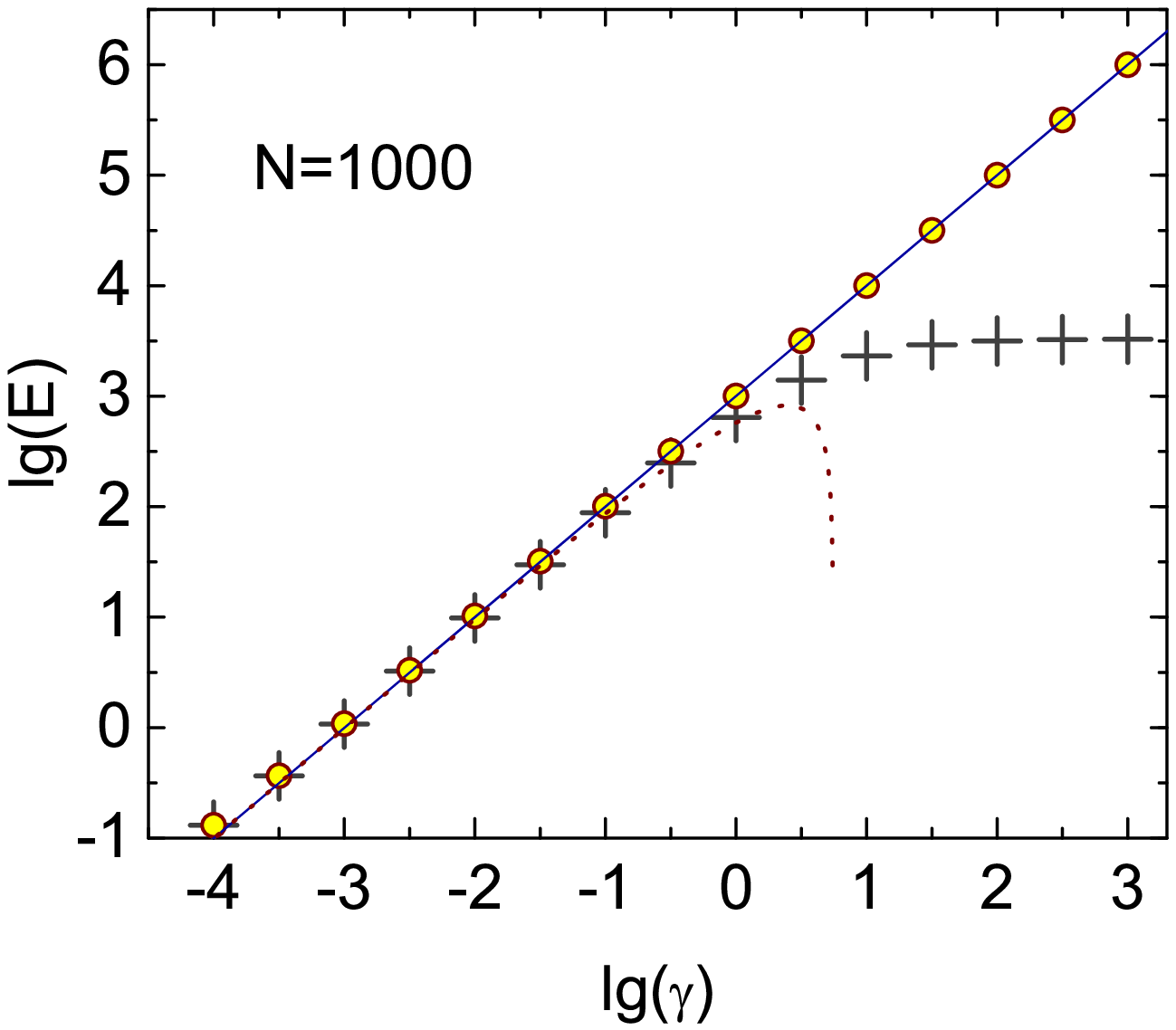}
\caption{ [Color online] The dependence of the ground state energy
$E$ on $\gamma $ calculated in various models for $\bar{\rho}=1$ and
$N=2$, $10$, $100$, $1000$: $E_{\mathrm{GP}}$ (circles),
$E_{\mathrm{GP_{N}}}$ (diamonds), the exact energy
$E_{\mathrm{Bethe}}$ (Eqs.~(\ref{5-2}) and (\ref{5-3}) with
$n_{j\leq N}=1$, crosses),  and
the Bogoliubov energy $E_{\mathrm{Bog}}=N \bar{\rho}^{2}\gamma \left( 1-\frac{4\sqrt{%
\gamma }}{3\pi }\right) $ (dotted curve). The solid lines in the
panels $N=2$, $10$, $100$ correspond to the ground state energy of
$N$ free particles, $E=N(\pi /L)^{2}$, and in the panel $N=1000$ to
the curve $E=N \bar{\rho}^{2}\gamma $. In the panels $N=2$ and $10$,
the circles are inside the diamonds at $\lg {\gamma }\leq -1$. In
the panels $N=100$ and $1000$, the curve $E_{\mathrm{GP_{N}}}(\gamma
)$ would be very close to $E_{\mathrm{GP}}(\gamma )$ and therefore
is not shown. } \label{fig3-6gseN}
\end{center}
\end{figure}

In the GP approach ($1-N^{-1}\rightarrow 1$), Eqs.~(\ref{14}) and
(\ref{norm3}) possess scaling properties: they do not change if
$\gamma $ and $N$ vary provided $\gamma
N^{2}=\mathrm{const}$. That is why the coefficients $g_{l}$ for such pairs $%
(\gamma ,N)$ are identical. The particle density profiles $\rho (x)$
are also the same for them if $\bar{\rho}$ are the same. However,
the energy $E_{GP}$ (\ref{5-0}) is different for each such $(\gamma
,N)$ pair.

\subsection{Ground state energy}

In this section, we find the energies $E_{\mathrm{GP}}$ and $E_{%
\mathrm{GP_{N}}}$ for the condensate ground state (formula (\ref{5-0}) with $j_{0}=1$, $%
\hbar =2m=1$) and analyse their dependences on $N$ and $\gamma $. We
also compare $E_{\mathrm{GP}}$ and $E_{\mathrm{GP_{N}}}$ with the
exact energy $E_{\mathrm{Bethe}}$ obtained from Eqs.~(\ref{5-2}),
(\ref{5-3}).

The dependence $E(\gamma )$  for $N=2$, $10$, $100$, $1000$ is shown
in Fig.~\ref{fig3-6gseN}. If $\gamma =0$, the ground state
corresponds to $N$ free particles with the total energy $E=N(\pi
/L)^{2}$.
As can be seen from the figures, the energies $E_{\mathrm{GP}}$ and $E_{%
\mathrm{GP_{N}}}$ are close to the energy of free particles if
$\gamma $ is small. The larger $N$, the smaller $\gamma $ at which
such a nearness of energy values takes place (because as $N$
increases, the potential energy increases faster than the kinetic
one). At $\gamma \rightarrow \infty $, the exact solution
$E_{\mathrm{Bethe}}$ tends to the limit of impenetrable bosons (for
$N\rightarrow \infty $, this is $E_{\mathrm{Bethe}}=N\pi
^{2}n^{2}/3$ \cite{girardeau1960}). Therefore, the curve $E_{\mathrm{Bethe}%
}(\gamma )$ saturates at $\gamma \gg 1$. In this case, the energies $E_{%
\mathrm{GP}}$ and $E_{\mathrm{GP_{N}}}$ differ strongly from the
exact energy.
However, if $\gamma \lesssim 0.1$, the energies $E_{\mathrm{GP}}$ and $E_{%
\mathrm{GP_{N}}}$ are close to the exact energy $E_{\mathrm{Bethe}}$
(we verified it for $\bar{\rho}=0.1$, $1$, and $10$). Moreover, for
$N\gtrsim 100$, the energies $E_{\mathrm{GP}}$ and
$E_{\mathrm{GP_{N}}}$ are close to the Bogoliubov ground state
energy if $\gamma $ is small but not too small.

The dependence $E(N)$ for different $\gamma $ is shown in Fig.~\ref%
{fig7-9gse}. From the dependence for $c=1$, one can see that the energy $E_{%
\mathrm{GP_{N}}}$ is close to $E_{\mathrm{Bethe}}$ even for a fairly
large $\gamma$ ($\gamma
=1$) if $N=2$ or $3$, but $E_{\mathrm{GP}}$ differs appreciably from $E_{%
\mathrm{Bethe}}$ for $\gamma =1$ and any $N$. For $\gamma =0.1$ the energies $%
E_{\mathrm{GP}}$ and $E_{\mathrm{GP_{N}}}$ are close to
$E_{\mathrm{Bethe}}$
if $N\lesssim 100$. Finally, for $\gamma =0.01$, the energies $E_{\mathrm{GP}%
}$, $E_{\mathrm{GP_{N}}}$ and $E_{\mathrm{Bethe}}$ are close to each
other for all $N$.

\begin{figure}
\begin{center}
\includegraphics[width=.32\textwidth]{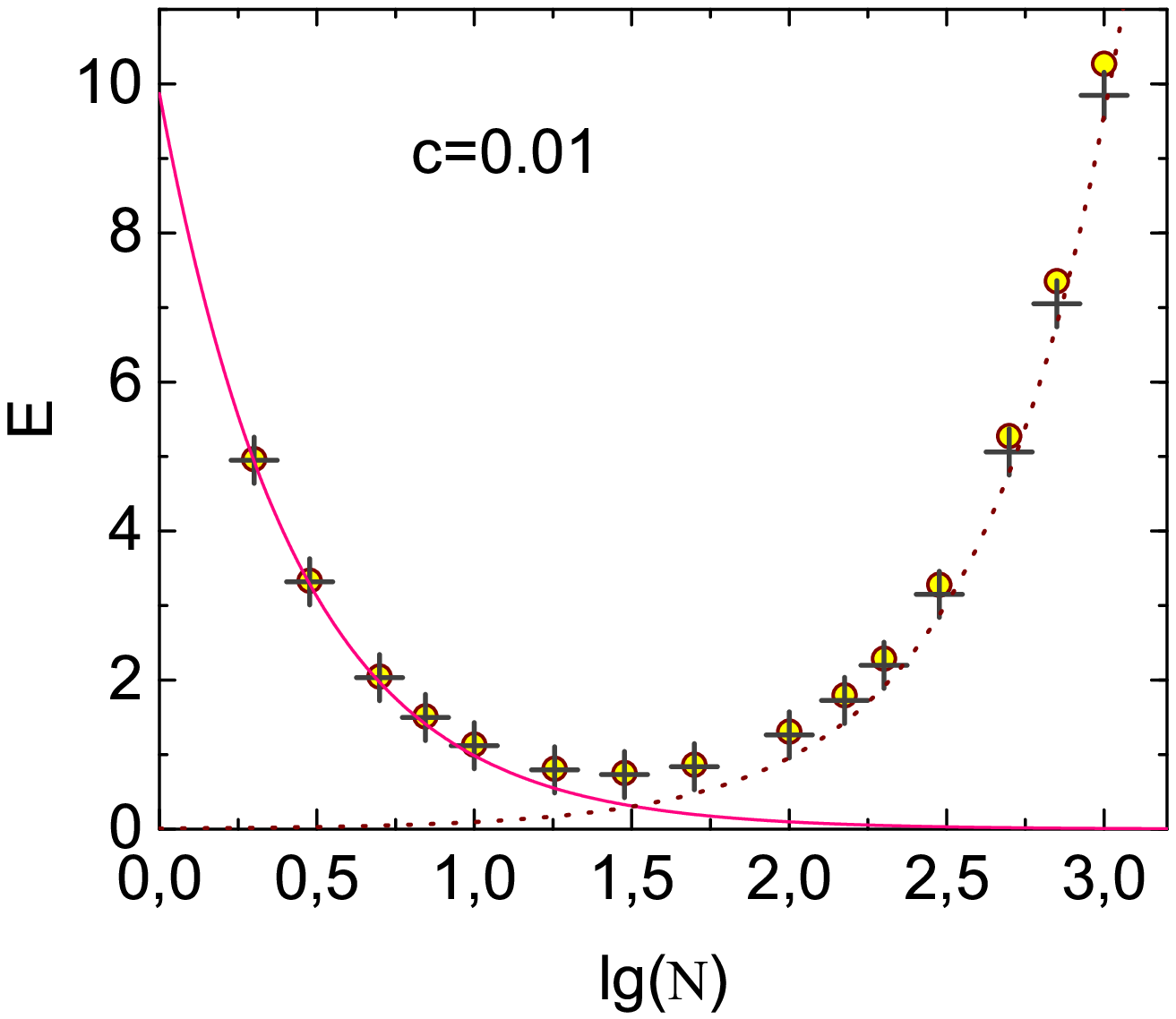}
\includegraphics[width=.32\textwidth]{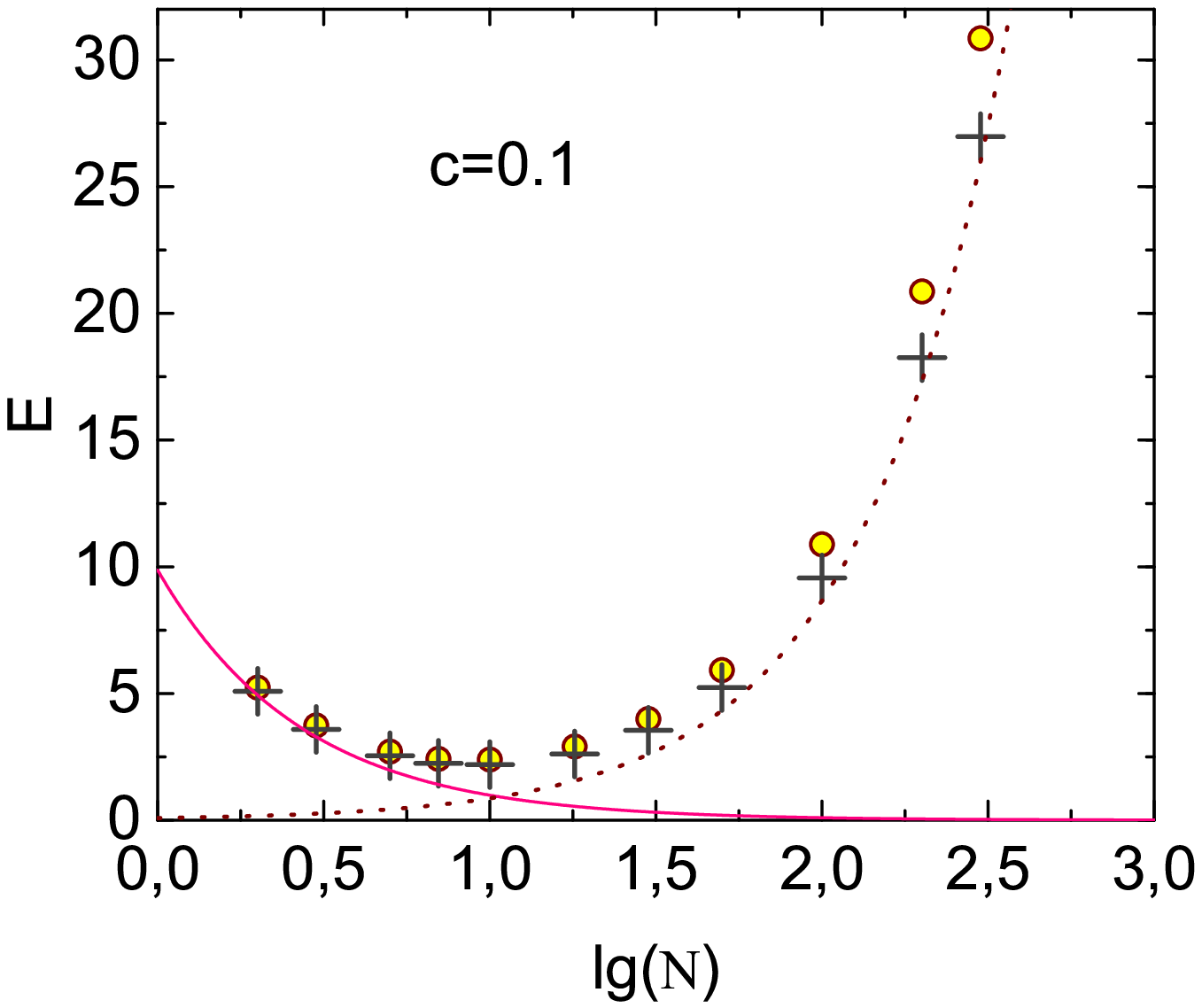}
\includegraphics[width=.32\textwidth]{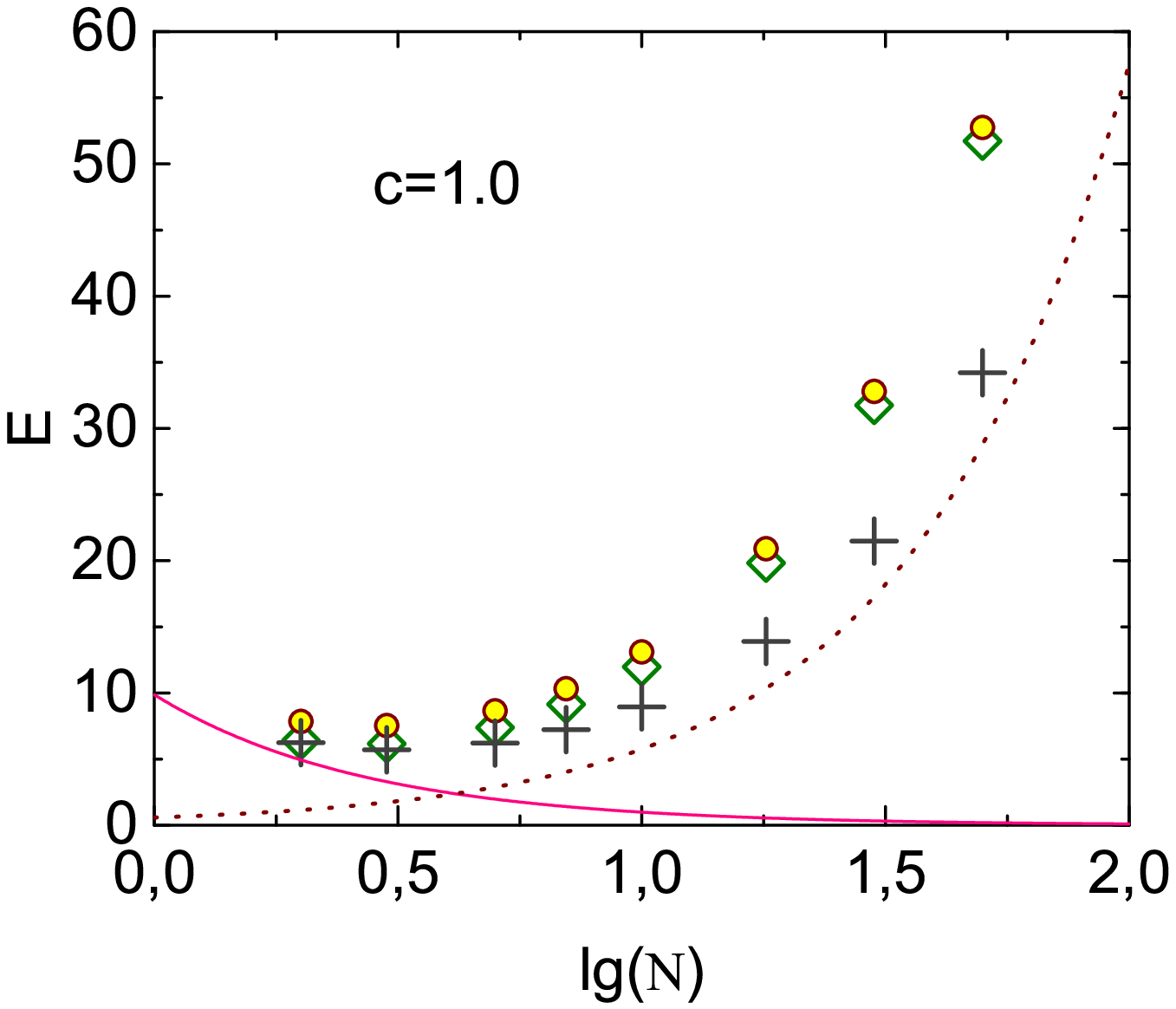}
\caption{ [Color online] The dependence of the ground state energy
$E$ on $N $ calculated in different models for $\bar{\rho}=1$ and
$c=0.01$, $0.1$, $1$: $E_{\mathrm{GP}}$ (circles),
$E_{\mathrm{GP_{N}}}$ (diamonds), the exact energy
$E_{\mathrm{Bethe}}$ (crosses),  the Bogoliubov energy
$E_{\mathrm{Bog}}=N \bar{\rho}^{2}\gamma \left( 1-
\frac{4\sqrt{\gamma }}{3\pi }\right)$ (dotted line), the ground
state energy of $N$ free particles $E=N(\pi /L)^{2}$ (solid line).
In the panels $c=0.01$ and $0.1$, the curve $E_{\mathrm{GP_{N}}}(N)$
is not shown because it is very close to $E_{\mathrm{GP}}(N)$. }
\label{fig7-9gse}
\end{center}
\end{figure}

\textit{For all cases shown in the figures and not shown, the energy
$E_{\mathrm{GP_{N}}}$ is closer to the exact energy
$E_{\mathrm{Bethe}}$ than the energy $E_{\mathrm{GP}}$, for all
values of the parameters.} In this case, for $N\gg 1$ the energies
$E_{\mathrm{GP}}$ and $E_{\mathrm{GP_{N}}}$ are very close to each
other, whereas for $N\lesssim 10 $ they are appreciably different,
with $E_{\mathrm{GP_{N}}}$ being much closer to
$E_{\mathrm{Bethe}}$.

Note the following interesting feature. One can see in panel $c=0.01$ in Fig.~\ref%
{fig7-9gse} that the condensate is close to the system of free
particles when $N\lesssim 10$, and to Bogoliubov's system when
$N\gtrsim 300$. For $N\sim 30\div 100$, the system has already left
the free particle regime but has not yet entered Bogoliubov's one;
in this case, the energies $E_{\mathrm{GP}}$ and
$E_{\mathrm{GP_{N}}}$ are close to the exact energy
$E_{\mathrm{Bethe}}$. That is, for $N\simeq 30$--$100$ structure
(\ref{3}) seems to work fairly well, and the interaction between the
particles manifests itself more in the change in the form of $\Phi
(x)$ rather than in interparticle correlations.

\subsection{Particle density profile for the ground state}

The local particle density for a 1D system of $N$ bosons in the
state $\Psi (x_{1},\ldots ,x_{N})$ is defined by the formula
\begin{equation}
\rho (x)=N\int dx_{2}\ldots dx_{N}|\Psi (x,x_{2},\ldots ,x_{N})|^{2}.
\label{d-ro1}
\end{equation}%
For $N$ free particles being in the ground state $\Psi (x_{1},\ldots
,x_{N})=\prod_{j=1,\ldots ,N}\sqrt{2/L}\sin {(\pi x_{j}/L)}$, it
gives
\begin{equation}
\rho (x)=2\bar{\rho}\sin ^{2}{(\pi x/L)},\quad \bar{\rho}=N/L.
\label{d-ro2}
\end{equation}%
In the GP$_{N}$ approach with the condensate ansatz (\ref{3}), we obtain $%
\rho (x)=N|\psi (x,t)|^{2}$ with the normalisation
$\int_{0}^{L}dx|\psi (x,t)|^{2}=1$.  Setting $\psi (x,t)=e^{\epsilon
t/i\hbar }e^{i\kappa(t)/\hbar}\Phi (x)/\sqrt{N}$, we get
\begin{equation}
\rho (x)=|\Phi (x)|^{2},  \label{d-ro3}
\end{equation}%
where $\Phi (x)$ is the wave function from the GP$_{N}$ equation (\ref{9-2}%
). The operator GP$_{N}$-approach also leads to (\ref{d-ro3}): $\hat{\Psi}(%
\mathbf{r},t)=\hat{a}_{0}\Psi (\mathbf{r},t)/\sqrt{N}$, $\rho
(x)=\langle \hat{\Psi}^{+}(x,t)\hat{\Psi}(x,t)\rangle _{T=0}=\langle
\hat{N}_{0}\rangle _{T=0}|\Psi (x,t)|^{2}/N=|\Psi (x,t)|^{2}=|\Phi
(x)|^{2}$ (here $\langle \hat{A}\rangle _{T=0}=\int dx_{1}\ldots
dx_{N}\Psi _{N}^{\ast }\hat{A}\Psi _{N}=\langle N|\hat{A}|N\rangle
$). Similarly, within the GP approach we find $\hat{\Psi}(x,t)=\Psi (x,t)$ and $\rho (x)=\langle \hat{%
\Psi}^{+}(x,t)\hat{\Psi}(x,t)\rangle _{T=0}=|\Psi (x,t)|^{2}=|\Phi (x)|^{2}$.

\begin{figure}
\includegraphics[width=0.47\textwidth]{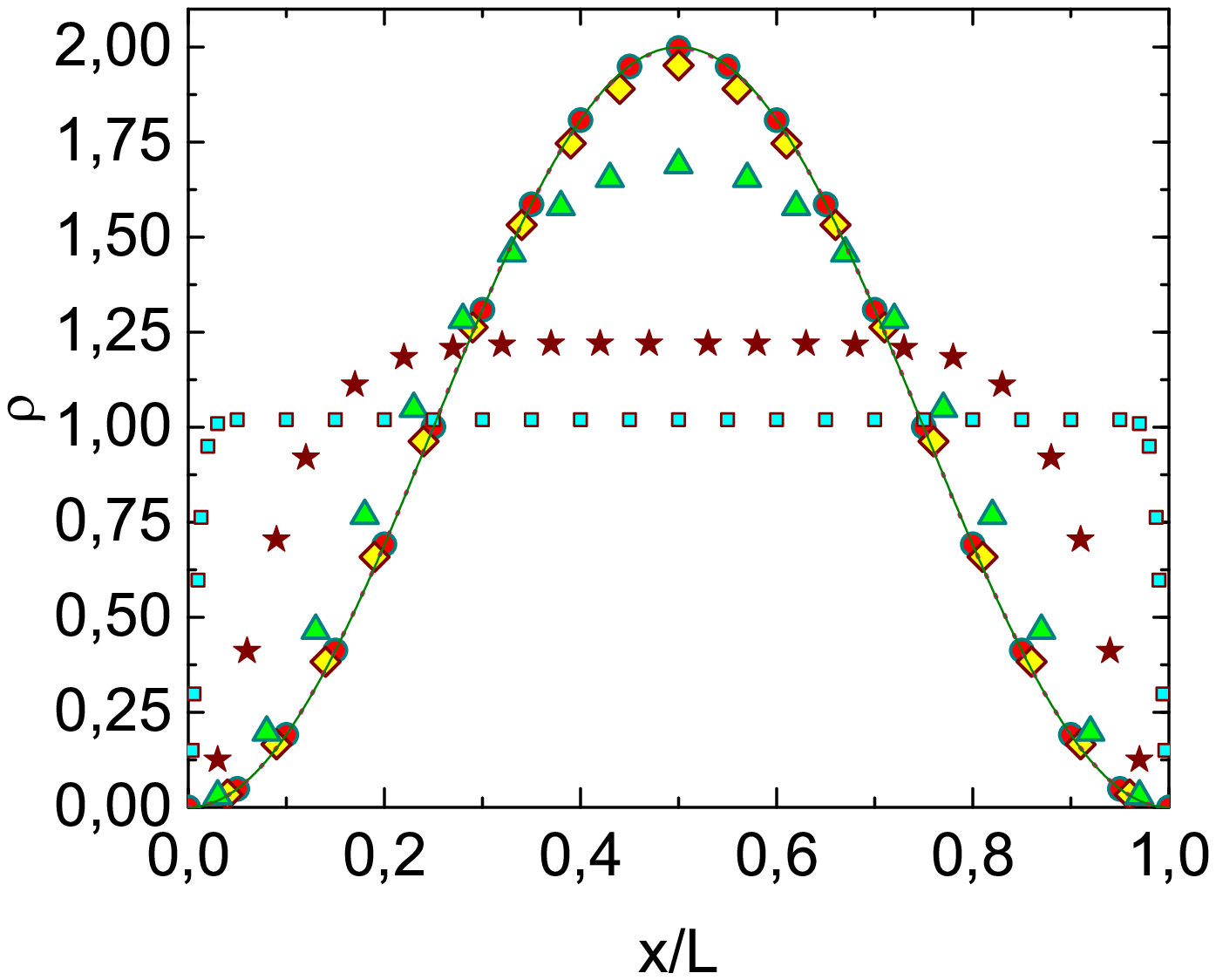}
\hfill
\includegraphics[width=0.47\textwidth]{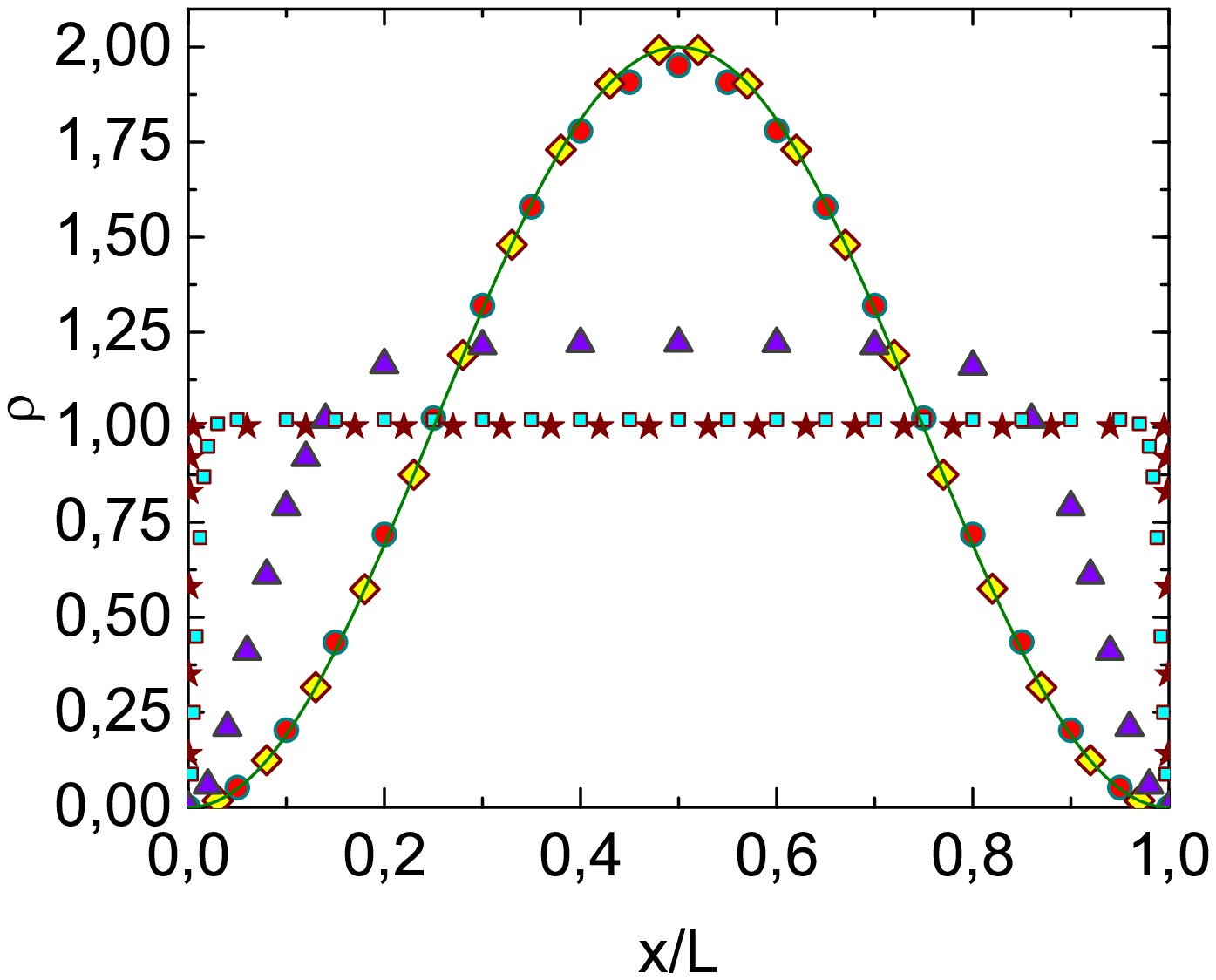}
\\
\parbox[t]{0.47\textwidth}{
\caption{ [Color online] The ground-state density profile $\rho (x)$
calculated for $\gamma=0.01$, $\bar{\rho}=1$, and various $N$:
$\rho(x)$, obtained from the GP$_{N}$ equation with $N=2$ (circles),
$10$ (diamonds), $ 30$ (triangles), $100$ (asterisks), $1000$
(squares); the exact Bethe-ansatz solution for $N=2$ (dotted curve).
The solid line marks  the density profile $\rho (x)=2\bar{\rho}\sin
^{2}{(\pi x/L)}$ for a system of any number ($N\geq 1$) of free
particles with $\bar{\rho}=1$. The solid and dotted curves are
almost indistinguishable from each other. \label{fig10gsro}} }
\hfill
\parbox[t]{0.47\textwidth}{
\caption{ [Color online] The ground-state density profile $\rho (x)$
calculated within the GP$_{N}$ approach for $N=1000$,
$\bar{\rho}=1$, and various $\gamma$:  $\gamma =10^{-8}$ (diamonds),
$10^{-6}$ (circles), $10^{-4}$ (triangles), $10^{-2}$ (squares), and
$1$ (asterisks). The solid curve is the  density profile $\rho
(x)=2\bar{\rho}\sin ^{2}{(\pi x/L)}$ for a system of free particles
with $\bar{\rho}=1$. \label{fig11gsro}} }
\end{figure}

Let us determine $\rho (x)$ for the GP and GP$_{N}$ approaches on the basis
of Eqs.~(\ref{11}), (\ref{14}), (\ref{norm3}), and (\ref{d-ro3}). The
results of numerical analysis are depicted in Figs.~\ref{fig10gsro} and~\ref%
{fig11gsro}. The GP and GP$_{N}$ curves $\rho (x)$ are close to each
other and would be visually indistinguishable. Therefore, only the
GP$_{N}$ curves are plotted.

Figure~\ref{fig10gsro} demonstrates the $\rho (x)$-profile for
$\gamma =0.01 $ and different $N$. The curves $\rho (x)$ for $N=10$
and $N=30$ are similar to the curve for $N=2$ and contain no
intervals with a constant particle density $\rho
(x)=\mathrm{const}$. Such $\rho (x)$ are close to
those for free particles. On the contrary, for $N\gtrsim 100,$ the profile $%
\rho (x)$ contains a large section where $\rho (x)=\mathrm{const}$;
it shows that the collective properties of the system manifest themselves at $%
N\gtrsim 100$. Note that the profiles $\rho (x)$ obtained in the GP and GP%
$_{N}$ approaches for $\gamma =0.01$, $\bar{\rho}=1$, $N=2$ coincide
with high accuracy with the profile $\rho (x)$ found by the Bethe
ansatz (see Fig.~\ref{fig10gsro}).

Figure~\ref{fig11gsro} shows the profiles $\rho (x)$ for $N=1000$
and different $\gamma $. Let us introduce the half-width $\eta $ of
the wall layer; this parameter is equal to the smallest coordinate
$x$ for which $\rho (x)=\rho (L/2)/2$. Numerical analysis has shown
that for $N=2$ and $\gamma \lesssim 1$, the quantity $\eta $ is
practically independent of $\gamma $ and is equal to $\eta \approx
L/4$. For $N=30$ and $0.1\lesssim \gamma
\lesssim 1$, $\eta $ considerably depends on $\gamma $. And for $N=1000$ and $%
N^{-2}\ll \gamma \lesssim 1$, we have $\eta \approx
\frac{1}{\bar{\rho}\sqrt{\gamma }} $ (a close estimate $\eta \approx
\frac{\pi }{2 \bar{\rho}\sqrt{\gamma }}$ was obtained in
\cite{mtmethodbog}). For $\gamma N^{2}\lesssim 1$ and any $N\geq 2
$, the interval with the constant particle density disappears, and
the system is in the near-free particle regime. In this case, the dependence $%
\rho (x)$ for $N$ interacting bosons is close to
dependence~(\ref{d-ro2}) for free bosons, and the concept of the
wall layer partly loses its meaning, although we may assume that
$\eta =L/4$.

\subsection{Near-free particle regime}

The regime of near-free particles corresponds to the condition
$\gamma N^{2}\lesssim 1$. For $N\gg 1$, this relation corresponds to
ultraweak coupling, $\gamma \lesssim N^{-2}$, and the ground state
energy is approximately described by the formula
\cite{batchelor2005,mt2015}
\begin{equation}
E_{af}\approx \frac{N\pi ^{2}}{L^{2}}+\frac{3(N-1)\gamma
\bar{\rho}^{2}}{2}.
\end{equation}%
For example, for $N=1000$, $\bar{\rho}=1$, and $\gamma =10^{-8}$, we have $%
E_{af}=0.0098845894\equiv N\pi ^{2}/L^{2}+0.000014985$, $E_{\mathrm{GP}%
}=0.0098846031\equiv N\pi ^{2}/L^{2}+0.0000149987$, and $E_{\mathrm{GP_{N}}%
}=0.0098845881376\equiv N\pi ^{2}/L^{2}+0.000014983738$. The exact solution
is $E_{\mathrm{Bethe}}=0.009884588128\equiv N\pi ^{2}/L^{2}+0.000014983728$.

\begin{figure}
\begin{center}
\includegraphics[width=.32\textwidth]{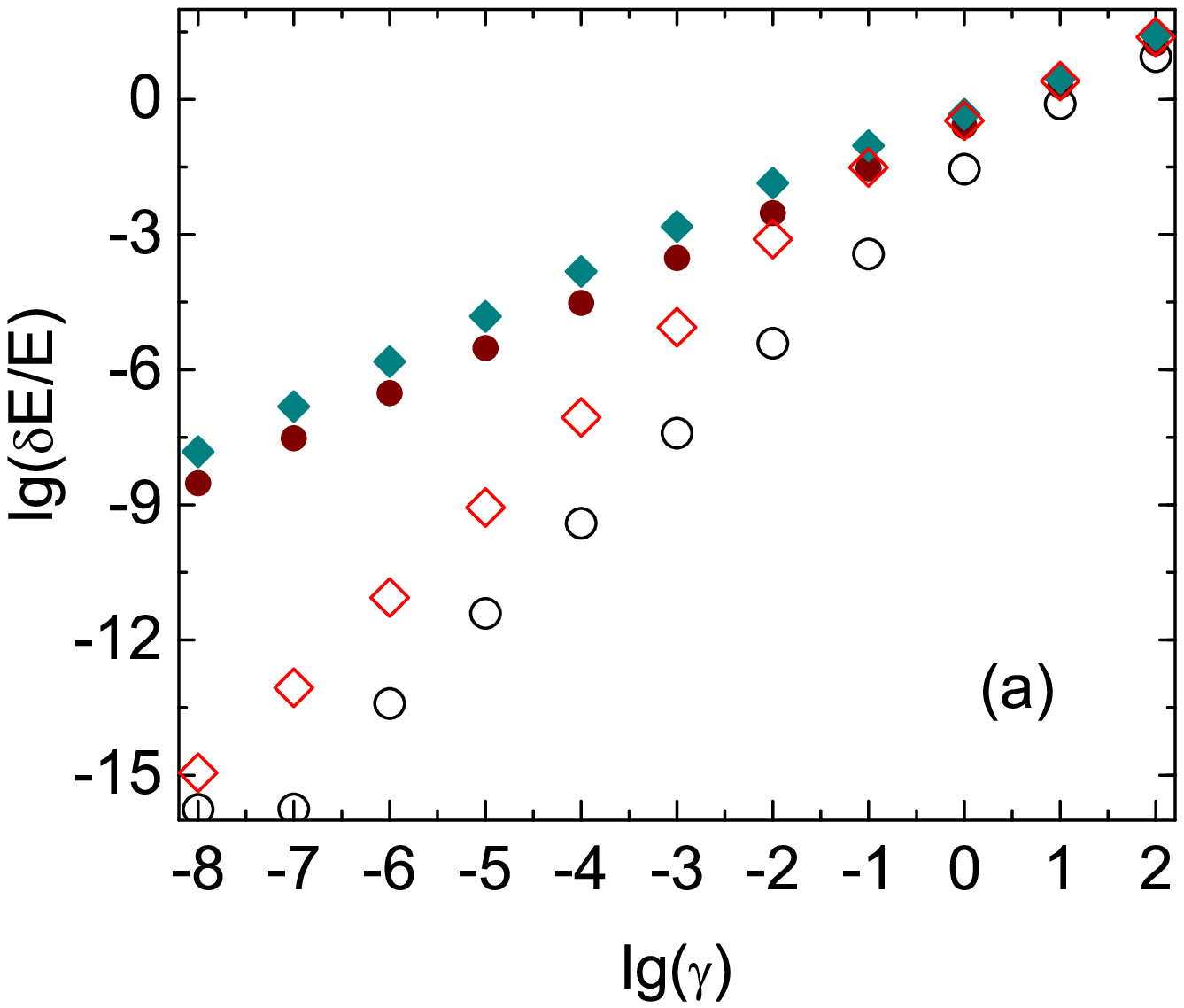}
\includegraphics[width=.32\textwidth]{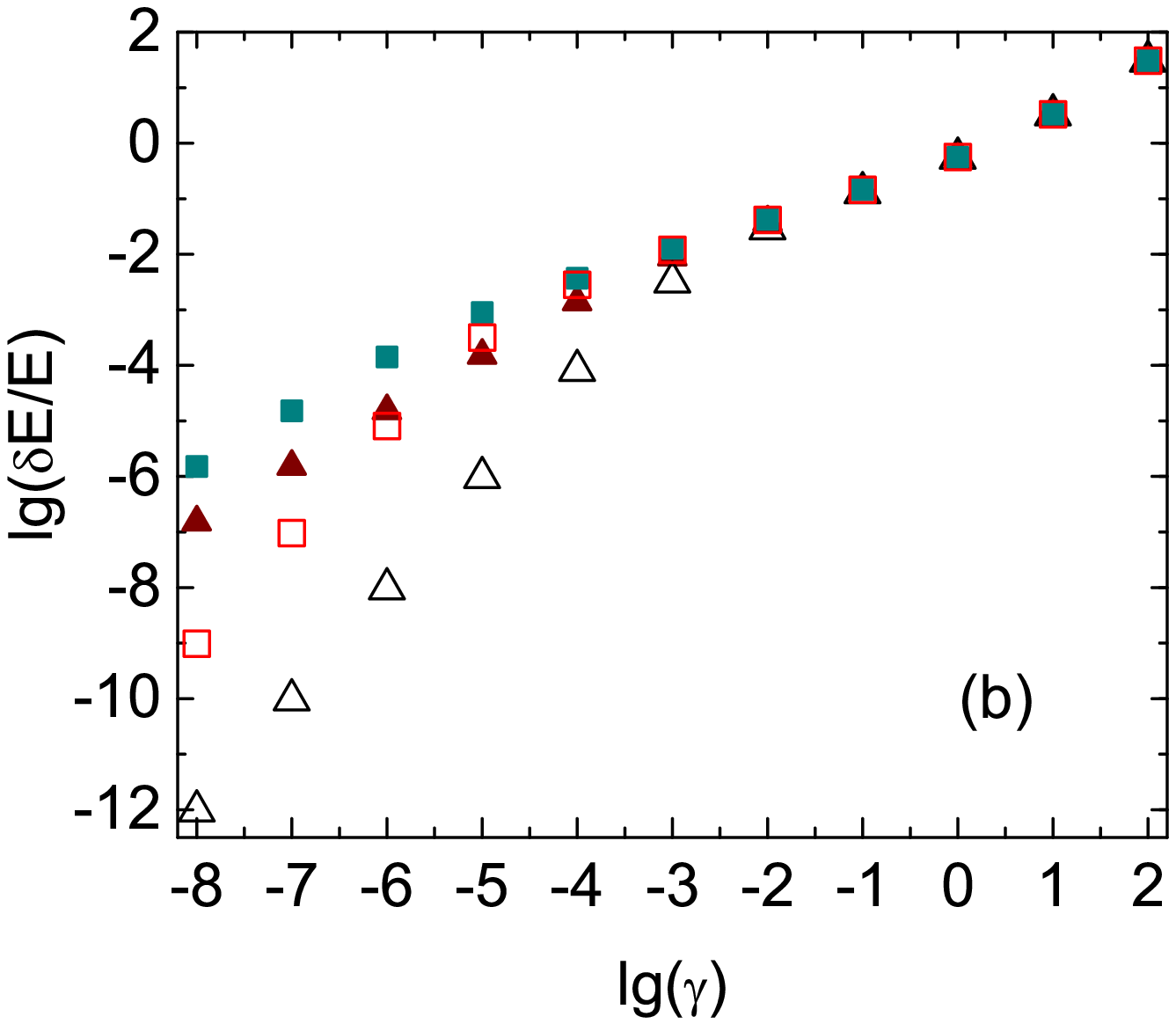}
\includegraphics[width=.32\textwidth]{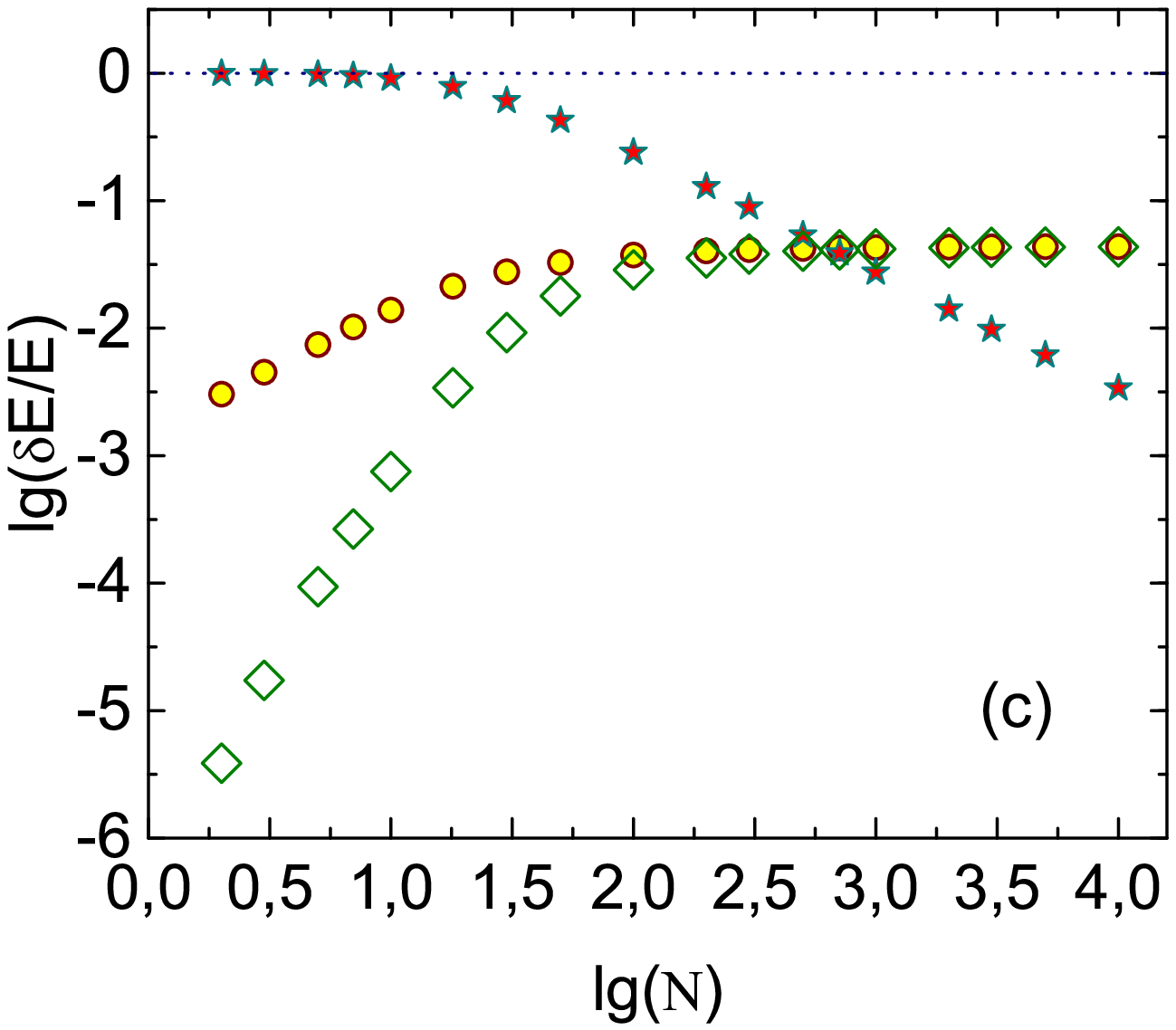}
\caption{ [Color online] Comparison of the accuracy of the GP and
GP$_{N}$ approaches for $\bar{\rho}=1$. \textbf{(a)} $\gamma
$-dependences of the ratio
$\frac{E_{\mathrm{GP}}-E_{\mathrm{Bethe}}}{E_{\mathrm{Bethe}}}$ for
$N=10$ (solid diamonds) and $N=2$  (solid circles), and the ratio
$\frac{E_{\mathrm{GP_{N}}}-E_{\mathrm{Bethe}}}{E_{\mathrm{Bethe}}}$
for $N=10$ (hollow diamonds) and $N=2$ (hollow circles).
\textbf{(b)} $\gamma $-dependences of the ratio $\frac{E_{\mathrm{GP}}-E_{\mathrm{Bethe}}}{E_{%
\mathrm{Bethe}}}$ for $N=100$ (solid triangles) and $N=1000$ (solid
squares), and the ratio
$\frac{E_{\mathrm{GP_{N}}}-E_{\mathrm{Bethe}}}{E_{\mathrm{Bethe}}}$
for $N=100$ (hollow triangles) and $N=1000$ (hollow squares).
\textbf{(c)}~Dependences of the ratios
$\frac{E_{\mathrm{GP}}-E_{\mathrm{Bethe}}}{E_{\mathrm{Bethe}}}$
(circles),
$\frac{E_{\mathrm{GP_{N}}}-E_{\mathrm{Bethe}}}{E_{\mathrm{Bethe}}}$
(diamonds),
and $\frac{E_{\mathrm{Bethe}}-E_{\mathrm{Bog}}}{E_{\mathrm{%
Bethe}}}$ (asterisks) on $N$ for $c=0.01$. Here $E_{\mathrm{Bog}%
}=N \bar{\rho}^{2}\gamma \left( 1-\frac{4\sqrt{\gamma }}{3\pi
}\right) $ is the Bogoliubov ground state energy of the 1D system of
$N$ point
bosons. The dotted line marks the zero level: $\lg {(\delta E/E)}=0$. }%
\label{fig12-14gs}
\end{center}
\end{figure}

In Fig.~\ref{fig12-14gs} the values of  $\frac{E_{\mathrm{GP}%
}-E_{\mathrm{Bethe}}}{E_{\mathrm{Bethe}}}$ and $\frac{E_{\mathrm{GP_{N}}}-E_{%
\mathrm{Bethe}}}{E_{\mathrm{Bethe}}}$ are plotted for $\bar{\rho}=1$ and different $\gamma $ and $%
N$. Since $E_{\mathrm{Bethe}}$ is the exact solution, \textit{Fig.~%
\ref{fig12-14gs} illustrates the accuracy of the GP and GP$_{N}$ approaches.}%
It is easy to see that for $\gamma N^{2}\lesssim 1$, i.e. in the
near-free particle regime, the GP$_{N}$ approach is in much better
agreement with the exact one than the GP approach.

\section{Comparison of the GP and GP$_{N}$ approaches}

Fig.~\ref{fig12-14gs} shows several patterns of relationships.
First, both
approaches reproduce the exact energy $E_{\mathrm{Bethe}}$ well for all $%
N\geq 2$ if $\gamma \lesssim 0.1$. Second, the GP and GP$_{N}$
approaches give similar results for large $N$. However, for small
$N$ and $\gamma N^{2}\lesssim 1$, the GP$_{N}$ solutions agree
\textit{much better}, than the GP ones, with the exact solutions.
For example, for $N=2$, $\bar{\rho}=1$, $\gamma =10^{-3}$, the
estimates $\frac{E_{\mathrm{GP}}-E_{\mathrm{Bethe}}}{E_{\mathrm{Bethe}}}%
\approx 10^{-3.5}\sim \gamma $ and $\frac{E_{\mathrm{GP_{N}}}-E_{\mathrm{Bethe}}%
}{E_{\mathrm{Bethe}}}\approx 10^{-7.4}\sim \gamma ^{2.5}$ hold, whereas for $%
N=2$, $\bar{\rho}=1$, $\gamma =10^{-7}$ we have $\frac{E_{\mathrm{GP}}-E_{%
\mathrm{Bethe}}}{E_{\mathrm{Bethe}}}\approx 10^{-7.5}\sim \gamma $ and $%
\frac{E_{\mathrm{GP_{N}}}-E_{\mathrm{Bethe}}}{E_{\mathrm{Bethe}}}\approx
10^{-15.7}\sim \gamma ^{2.2}$. In the latter case, $E_{\mathrm{GP}%
}=4.9348025$, $E_{\mathrm{GP_{N}}}=4.9348023505446781$, and $E_{\mathrm{Bethe}%
}=4.9348023505446772$. That is, for $\gamma =10^{-7}$ the relative
error of the GP$_{N}$ solution is about $10^{-16}$. This is amazing
accuracy!

A more general property holds: the GP$_{N}$ approach works much
better
than the GP one in the near-free particle regime ($\gamma N^{2}\lesssim 1$%
), which corresponds to small $N$ ($N\lesssim \gamma ^{-1/2}$) or small $%
\gamma $ ($\gamma \lesssim N^{-2}$). Why? In our opinion, this is a
result of the following: It is natural to expect that, with a high
probability, every near-free boson is in the condensate, for any
$N\geq 2$. Therefore, ans\"{a}tze (\ref{3}), (\ref{5}), and
$\hat{\Psi}(x,t)=\Psi (x,t)$ are good approximations. Since
ans\"{a}tze (\ref{3}) and (\ref{5}) are
more precise than the c-number ansatz $\hat{\Psi}(x,t)=\Psi (x,t)$, the GP$%
_{N}$ approach turns out to be more accurate than the GP one.

It is commonly believed that the GP approach works only for large
$N$ because the approximation $\hat{\Psi}(x,t)=\Psi (x,t)$ is
reasonable only when $N\gg 1$. In point of fact, the GP approach
works even \textit{better} for small $N$ than for large $N$ (see
Fig.~\ref{fig12-14gs}). This surprising property appears to be due
to the fact that the GP equation is simply close to the GP$_{N}$
one, which describes few-particle systems with high accuracy.

In turn, the following question arises: Why is the GP$_{N}$ equation
works better in the case of small $N$? The evident answer is that
the role of two- and many-particle correlations is less for small
$N$ (such correlations are not taken into account in ans\"{a}tze
(\ref{3}), (\ref{5})). More information can be obtained from the
diagonal expansion of the single-particle density matrix
\begin{equation}
F_{1}(x,x^{\prime })=\sum \limits_{j=1}^{\infty }\lambda _{j}\phi _{j}^{\ast
}(x^{\prime })\phi _{j}(x),  \label{6-1}
\end{equation}%
where $\lambda _{j}$ are the occupation numbers of the
single-particle states $\phi _{j}(x)$ (in this case, $\lambda
_{1}+\lambda _{2}+\ldots +\lambda _{\infty }=N$), and $\phi _{j}(x)$
form the complete collection of orthonormal functions. Condensate
ans\"{a}tze (\ref{3}) and (\ref{5}) describe
the system well only if $\sqrt{N}\phi _{1}(x)$ is close to the solution $%
\Phi (x)$ of the GP$_{N}$ equation, and the relations $\lambda
_{1}\simeq N$, $\lambda _{j>1}\ll N$ hold. We suppose that
$(\frac{\lambda_{1}(\gamma)}{N}|_{N\leq 10})>(\frac{\lambda
_{1}(\gamma)}{N}|_{N\geq 100})$ for $\gamma \lesssim 0.1$. If so,
then the role of two- and many-particle correlations is less for
$N\lesssim 10$ than for $N\gtrsim 100$.

Next, one can see from Fig.~\ref{fig12-14gs}c that the Bogoliubov
energy $E_{\mathrm{Bog}}$ for large $N$ is closer to the exact energy $E_{%
\mathrm{Bethe}}$ than the energies $E_{\mathrm{GP}}$ and $E_{\mathrm{GP_{N}}}$%
. That is, the Bogoliubov approach describes the ground state of the
system with large $N$ more accurately than the GP and GP$_{N}$
approaches. This property shows that taking into account
above-condensate atoms---which is done in the Bogoliubov model, but
not in the GP and GP$_{N}$ approaches---is significant.

In order to better understand the properties of the system, it is
necessary to find the  density matrix (\ref{6-1}) and an equation
for the condensate in the approximation that accurately takes into
account the above-condensate atoms (or, equivalently, two- and
many-particle correlations).

\section{Concluding remarks}

We have shown that the condensate ansatz
$\hat{\Psi}(\mathbf{r},t)=\hat{a}_{0}\Psi (\mathbf{r},t)/\sqrt{N}$
(\ref{5}) leads to a modified Gross-Pitaevskii equation  (the
GP$_{N}$ equation), which differs from the standard Gross-Pitaevskii
(GP) equation by the additional factor $\left( 1-1/N\right) $.  We
have analysed the ground state of the system for the mean particle
density $\bar{\rho}=1$ and compared the solutions of the GP and
GP$_{N}$ equations  with the exact solutions.  Our results evidence
that both equations describe well a system of $N\geq 2$ interacting
spinless bosons,  if the coupling is weak: $\gamma \lesssim 0.1$.
This significantly extends the conventional view that the
Gross-Pitaevskii equation only holds when $N\gg 1$.

In the case of the near-free particle regime, $\gamma N^{2}\lesssim
1$, the GP$_{N}$ equation describes the Bose system much better than
the standard Gross-Pitaevskii equation. The condition $\gamma
N^{2}\lesssim 1$ can be written as $%
\gamma \lesssim N^{-2}$ or $N\lesssim \gamma ^{-1/2}$, which
corresponds to ultraweak coupling or small $N$, respectively. Thus,
the Gross-Pitaevskii equation with the additional factor $\left(
1-1/N\right) $ accurately describes a Bose system with a small
number of particles, $N\lesssim \gamma ^{-1/2}$.

Since the GP$_N$ equation describes a few-boson system with high
accuracy, \textit{it is worth extending the concept of the
condensate to systems with small $N$}, using the following
criterion: if $\lambda_{1}\gg\lambda_{2}+\lambda_{3}+\ldots +
\lambda_{\infty}$
in (\ref{6-1}), then state~1 is a condensate state, whereas other
states are not. For an ideal gas, all $\lambda_{j}$ are integers,
and this criterion gives $\lambda_{1}=N;
\lambda_{2}=\lambda_{3}=\ldots = \lambda_{\infty}=0$.

Interestingly, the GP and GP$_{N}$ equations work well for all $N$ and $%
\gamma \lesssim 0.1$, although they completely ignore two-particle
and higher-order correlations. Such properties mean that these
correlations are weak when $\gamma \lesssim 0.1$.

To describe a system of $N$ spinless interacting bosons, several
methods have been proposed:
analytical methods for systems with $N\gg 1$ and weak coupling \cite%
{bog1947,yuv1,fey1954,bz1955,brueck1959,yuv2,pash2004}; numerical
methods for systems with $N=2\div 10$ and any, weak or strong,
coupling (such methods use various expansions in basis functions) \cite%
{blume2012,mch,zinner2016,sowinski2019}; and Monte Carlo methods for
systems with $N\sim 10\div 100$ and arbitrary
coupling~\cite{ceperley1992,whitlock2006}. Moreover, the exactly
solvable approach describes 1D systems of $N\geq 2$ point bosons,
for arbitrary coupling~\cite{gaudin1971,ll1963,lieb1963,mtsp2019}.
The GP and GP$_{N}$ equations provide another analytical method
for systems with $N\geq 2$ (in this case, the GP$%
_{N}$ equation describes systems with $2\leq N\lesssim 100$ more
accurately than the GP equation).  To our knowledge, the ground
state of Bose systems with $10\lesssim N\lesssim 100$ can be
accurately described only by the Monte Carlo method and the GP$_{N}$
approach.

\section*{Acknowledgments}

This research was supported by the National Academy of Sciences of
Ukraine (project No.~0121U109612) and the Simons Foundation (grant
No.~1030283).

\appendix

\section*{Appendix: Proof that $\Phi (x)$ is real}

Let us show that any solution $\Phi (x)$ of Eq.~(\ref{9-1}) with zero BCs (%
\ref{bc2}) can be written in a real form.

Let a solution of Eq.~(\ref{9-1}) be complex, $\Phi (x)=\Phi _{1}(x)+i\Phi
_{2}(x)$, where $\Phi _{1}(x)$ and $\Phi _{2}(x)$ are real functions. Then
Eq.~(\ref{9-1}) can be written as two equations:
\begin{equation}
\epsilon \Phi _{1}=-\frac{\hbar ^{2}}{2m}\frac{\partial ^{2}\Phi _{1}}{%
\partial x^{2}}+2c(\Phi _{1}^{2}+\Phi _{2}^{2})\Phi _{1},  \label{9-1a}
\end{equation}%
\begin{equation}
\epsilon \Phi _{2}=-\frac{\hbar ^{2}}{2m}\frac{\partial ^{2}\Phi _{2}}{%
\partial x^{2}}+2c(\Phi _{1}^{2}+\Phi _{2}^{2})\Phi _{2}.  \label{9-1b}
\end{equation}%
Multiplying Eq.~(\ref{9-1a}) by $\Phi _{2}(x)$ and Eq.~(\ref{9-1b}) by $\Phi
_{1}(x)$, and subtracting the results, we obtain the equation
\begin{equation}
\Phi _{2}\frac{\partial ^{2}\Phi _{1}}{\partial x^{2}}=\Phi _{1}\frac{%
\partial ^{2}\Phi _{2}}{\partial x^{2}}.  \label{p2-3}
\end{equation}%
Let us expand $\Phi _{1}(x)$ and $\Phi _{2}(x)$ in sine series,
\begin{equation}
\Phi _{1}(x)=\sum\limits_{j=1,2,\ldots ,\infty }a_{j}\sqrt{2/L}\cdot \sin
(k_{j}x),\quad \Phi _{2}(x)=\sum\limits_{j=1,2,\ldots ,\infty }b_{j}\sqrt{2/L%
}\cdot \sin (k_{j}x),  \label{p2-4}
\end{equation}%
where $k_{j}=\pi j/L$. Then Eq.~(\ref{p2-3}) can be written in the form
\begin{equation}
\sum\limits_{j_{1}j_{2}=1,2,\ldots ,\infty
}k_{j_{1}}^{2}(b_{j_{1}}a_{j_{2}}-b_{j_{2}}a_{j_{1}})\sin (k_{j_{1}}x)\sin
(k_{j_{2}}x)=0,  \label{p2-5}
\end{equation}%
or
\begin{equation}
(1/2)\sum\limits_{j_{1}j_{2}=1,2,\ldots ,\infty
}k_{j_{1}}^{2}(b_{j_{1}}a_{j_{2}}-b_{j_{2}}a_{j_{1}})[\cos
(k_{j_{1}-j_{2}}x)-\cos (k_{j_{1}+j_{2}}x)]=0.  \label{p2-6}
\end{equation}%
Using the formulae
\begin{equation}
\cos (k_{j}x)=\sum\limits_{p=1,2,\ldots ,\infty }c_{j}^{p}\sin (k_{p}x),
\label{p2-7}
\end{equation}%
\begin{equation}
c_{j}^{p}=\left[
\begin{array}{ccc}
0 & \ \mbox{for even}\ p-j, &  \\
\frac{2}{\pi }\left( \frac{1}{p-j}+\frac{1}{p+j}\right)  & \mbox{for
odd}\ p-j, &
\end{array}%
\right.   \label{p2-8}
\end{equation}%
we write Eq.~(\ref{p2-6}) as follows:
\begin{equation}
(1/2)\sum\limits_{j_{1}j_{2}=1,2,\ldots ,\infty
}k_{j_{1}}^{2}(b_{j_{1}}a_{j_{2}}-b_{j_{2}}a_{j_{1}})\sum\limits_{p=1,2,%
\ldots ,\infty }(c_{j_{1}-j_{2}}^{p}-c_{j_{1}+j_{2}}^{p})\sin (k_{p}x)=0.
\label{p2-9}
\end{equation}%
Let the function $\Phi _{1}(x)$ be known, and $\Phi _{2}(x)$ unknown. Since
the functions $\sin (k_{p}x)$ are independent,  from Eq.~(\ref{p2-9}%
) we obtain the system of equations for the unknown coefficients
$b_{j}$:
\begin{equation}
\sum\limits_{j_{1}j_{2}=1,2,\ldots ,\infty
}k_{j_{1}}^{2}(b_{j_{1}}a_{j_{2}}-b_{j_{2}}a_{j_{1}})(c_{j_{1}-j_{2}}^{p}-c_{j_{1}+j_{2}}^{p})=0,\quad p=1,2,\ldots ,\infty .
\label{p2-10}
\end{equation}%
Such a system of linear homogeneous equations for $b_{j}$, $%
\sum_{j}A_{pj}b_{j}=0$, always has a zero solution: $b_{j}=0$ for all $%
j=1,2,\ldots ,\infty $. If the determinant of the matrix $A_{pj}$ is
zero, then this system of equations also has an infinite number of
nonzero solutions $\{qb_{j}\}$, which differ from each other only by
the values of the general multiplier $q$. It is easy to see that the
quantities $b_{j}=ra_{j}$, where $r$ is the same for all
$j=1,2,\ldots ,\infty $, provide a solution of the
system~(\ref{p2-10}). The normalisation condition (\ref{norm}) leads
to the equation
\begin{equation}
\sum\limits_{j=1,2,\ldots ,\infty }(a^{2}_{j}+b^{2}_{j})=N,
\label{p2-11}
\end{equation}%
which allows one to find the value of $r$.  So, we obtain two
possible solutions: $\Phi _{2}(x)=r\Phi _{1}(x)$ and $\Phi
_{2}(x)=0$. In both cases, $\Phi (x)=\Phi _{1}(x)+i\Phi
_{2}(x)=\mathrm{const}\cdot \Phi _{1}(x)$. Therefore, we can
consider the function $\Phi (x)$ in Eq.~(\ref{9-1}) to be real. This
analysis can be easily generalized to two- and three-dimensional
cases.

\end{document}